\documentclass[12pt]{article}
 \usepackage{amsmath,amssymb}
 \newcommand{\be}{\begin{equation}}
 \newcommand{\ee}{\end{equation}}
 \newcommand{\bl}{\begin{equation}\begin{array}{ll}}
 \newcommand{\el}{\end{array}\end{equation}}
 \newcommand{\bll}{\begin{equation}\begin{array}{lll}}
 \newcommand{\bdm}{\begin{displaymath}}
 \newcommand{\edm}{\end{displaymath}}
 \def\bea{\begin{eqnarray}}
 \def\eea{\end{eqnarray}}
 \def\barr{\begin{array}}
 \def\earr{\end{array}}
 \newcommand{\bean}{\begin{eqnarray}}
 \newcommand{\eean}{\end{eqnarray}}
\def\p{\partial}

\def\f{\varphi}
\def\ve{\varepsilon}

 \def\De{\Delta}
 \def\de{\delta}
 \def\la{\lambda}
 \def\La{\Lambda}
 \def\al{\alpha}
 \def\ga{\gamma}
 \def\Ga{\Gamma}
 \def\ka{\kappa}
 \def\om{\omega}
 
 \def\sig{\sigma}
 \def\Sig{\Sigma}
\def\half{\frac{1}{2}}

\def\2third{\frac{2}{3}}
\def\4third{\frac{4}{3}}
\def\3quart{\frac{3}{4}}

\def\pr{\prime}

\def\bal{\bar{\alpha}}

\def\bv{\bar{v}}
\def\bl{\bar{l}}

\def\bk{\bar{k}}
\def\bl{\bar{l}}

\def\bchi{\bar{\chi}}

\def\cL{{\cal L}}

\def\da{\dot{a}}

\def\db{\dot{b}}

\def\drho{\dot{\rho}}

\def\dvphi{\dot{\f}}

\def\dxi{\dot{\xi}}

\def\dsig{\dot{\sigma}}
\def\dpsi{\dot{\psi}}
\def\dal{\dot{\alpha}}

\def\dbe{\dot{\beta}}

\def\dga{\dot{\gamma}}

\def\dA{\dot{A}}
\def\ddA{\ddot{A}}

\def\df{\dot{f}}
\def\dh{\dot{h}}

\def\hchi{\hat{\chi}}

\def\hchi{\hat{\chi}}

\def\Lag{{\cal L}}

\def\meqdef{\stackrel{\textbf {def}}{=}}

\def\prbe{\beta^{\prime}}
\def\prga{\gamma^{\prime}}
\def\prpsi{\psi^{\prime}}

\def\prbl{\bl^{\prime}}

\def\prchi{\chi^{\prime}}
\def\prchi2{(\chi^2)^{\prime\prime}}

\def\pra{a^{\prime}}

\def\pp{\prime\prime}

\def\tchi{\tilde{\chi}}
\def\tThe{\tilde{\Theta}}
\def\kbl{\textbf{k}}
\def\rbl{\textbf{r}}
\def\wbl{\textbf{w}}

\usepackage{amsmath}

\begin{document}

 \title{{\bf Constructing and solving cosmologies of early universes with dark energy and matter. I}}

 \author{Alexandre T.~Filippov \thanks{filippov@theor.jinr.ru} \\{\small \it {$^+$ Joint Institute for Nuclear Research, Dubna, Moscow Region RU-141980} }}
 \maketitle

 \begin{abstract}
   The dynamics of homogeneous spherical cosmologies with a scalar  or vector  massive particle, \emph{scalaron} or \emph{vecton}, 
   is described by three ordinary differential equations for two metric functions and the matter field depending on an  arbitrary evolution (`\emph{time}') parameter $\tau$. Unlike the standard practice, flatness and isotropy are assumed to be  \emph{emerging} properties of inflationary expansion and we propose for early cosmologies an  essentially anisotropic \emph{minimal affine extension} of general relativity, in which the nonmetric part of the connection is a vecton \emph{dark matter}.
   To understand this non-integrable theory we first formulate three simpler  models: linearized vecton cosmology and its scalaron analogs including the standard version. In the isotropic limit, the last is exactly solved by using 
   the metric $\al\sim\log\,(g_{rr})$ instead of $\tau$. The solution explicitly depends on a fairly arbitrary potential $\bv(\al)>0$ that can be explicitly related to the Hubble function $H(\al)$. The most general restrictions, $\bv(\al)>0$ and $\bv^\pr(\al)<0$, arise from \emph{positivity of the solutions}, which are canonical momenta squared. Other restrictions are related to cosmological scenarios, and it is possible to derive  potentials providing different scenarios, e.g., inflationary expansion. These are defined by \emph{three characteristic functions}: $0<\rbl(\al)<\infty$, $0<\wbl(\al)<1$, and $\Theta(\al)$. The first two  must be small and almost equal in the inflationary interval but grow near the exit, $\rbl\rightarrow +\infty$ if $\wbl\rightarrow 1$; $\Theta$ is positive before the exit from inflation. Within the framework of the model, we propose possible scenarios for the \emph{exit}: both functions reach a common maximum $\wbl_m$, $\rbl_m=\wbl_m/(1-\wbl_m)\gg 1$, after which they fast fall to the pre-inflationary values or below. The potential must very slowly grow (is almost constant) during inflation. Having \emph{exact analytic relations} between the potential $\bv$ (or the Hubble function) and basic functions $\rbl,\,\wbl$ it is natural to start constructing the models of early cosmologies with simple guesses for $\wbl$. This is most obvious, in case of inflationary models in which $\rbl,\,\wbl$ are very small during inflation and fast growing at its end. Such a `\emph{constructive' approach} looks promising also in vecton cosmology.\footnote{The author supposes that the reader is familiar with basics of general relativity \cite{Dau} (supplemented by some familiarity with Ch.1-2 of \cite{LPE}) and with modern cosmology,  including inflationary models as described in, e.g., \cite{Mukhanov}-\cite{Martin18}). In addition, the present paper is not completely independent of papers \cite{ATFnew}-\cite{ATFgen} and some knowledge of them will be helpful (they also contain many references to other  related work). Familiarity with original papers and recent reviews of the founders of the cosmological inflation theory A.Guth, A.Starobinsky, A.Linde is also desirable, although this theory is not the main subject of the present paper.}

   \end{abstract}

 \newpage
 \tableofcontents

 \section{Introduction}
 
 This paper is the beginning of reanimation and revision of my project started in 2008 \cite {ATF}, which proposed applying to modern cosmology ideas of H.Weyl, A.Einstein and A.Eddington on affine extension of general relativity that would naturally include electromagnetic field.\footnote{See a final discussion in \cite{Ein2}; the Lagrangian resembling our (\ref{e1}) was proposed in \cite{Einstein1}. These attempts failed on account of imaginary mass of the "photon", that signals "instability of the Universe" (Einstein).} In paper \cite {ATF} the mass problem was not seriously discussed and only a linearized kinetic energy was considered. More general nonlinear models including a massive vectons were  introduced in Refs.\cite{ATFn}-\cite{ATFg}. In particular, the nonlinear model discussed in \cite{ATFg} is now revised and proposed as a \emph{realistic unified model of dark matter and dark energy}. It is based on the minimal affine extension of general relativity preserving (up to parameterisations) the geodesics of the pure metric theory. It predicts a new massive vecton interacting only with gravity. The vecton dynamics is defined by a geometrically motivated mass term plus nonlinear kinetic energy term related to one proposed by Eddington and Einstein.\footnote{Nowadays it is commonly called the "DBI term", but we avoid this name related to electrodynamics.} For small vecton kinetic energy it is approximately the sum of the standard vector meson Lagrangian and the cosmological constant term. The final goal of the project is to solve this model in the low-energy and high-energy regimes presumably describing inflation and exit of it. To succeed, it is necessary to completely understand three simpler models: general scalaron model, linearized vecton model, and a special scalar theory with a non-linear kinetic term. The last object must imitate high energy behavior of the main nonlinear vecton theory.

 The foundation of this work lies on the new approach to scalaron dynamics  that allows to exactly solve the standard FLRW-type  scalaron cosmologies in inflationary regime (see \cite{ATFnew}-\cite{ATFgen}). It is important that this regime may in general emerge from non-flat, anisotropic initial state and that our approach is applicable to intrinsically anisotropic vecton cosmologies. Our mathematical formalism is transparent and makes it easy to find simple criteria for inflationary (and other) regimes. In fact, we can completely solve (algebraically and graphically) the \emph{inverse} problem -- to find \emph{potentials} providing early cosmologies with pre-assumed properties, which are defined in terms of three `characteristic' functions. Two of them are by definition positive, with derivatives that are positive up to the end of inflation.

 Moreover, we prepare the basis for applying this approach to much more complex anisotropic inflationary models of generalized gravity, dark energy and dark matter proposed in \cite{ATFn}-\cite{ATFg}. The main subject is a detailed analysis of general properties of the homogeneous scalaron cosmologies with asymptotically vanishing anisotropy. In particular, simple sufficient conditions defining potentials providing inflationary behavior will be formulated.
 
 With this goal, we first summarize one\,-\,dimensional cosmological reductions. They are supposed to give attractors that may correspond to possible early cosmologies. Chapter~3 presents and discusses the dynamical equations for all cosmologies which can be derived in this way. One of the keys for their solution is their Hamiltonian formulation. The transformation to our $\al$-representation  and the convenient gauge-invariant form of them are given in Chapter~4. The general  solution depending on the potential $\bv(\al)$ is explicit for vanishing anisotropy. Otherwise, for a wide class of the potentials, it can be derived iteratively in the frame of the evident \emph{asymptotic hierarchy}: $(\emph{flat, isotropic}) \subset \emph{isotropic} \subset \emph{anysotropic}$. This hierarchy, together with some other not so evident conditions, impose strong restrictions on the potential behavior in the domain $\al>0$.

 In fact, basing on consistency conditions we suppose that $\bv(\al)$ belongs to the ring of rational functions, postponing stronger restrictions to Chapter~5. In the rest of Chapter~4 we define and discuss important characteristic cosmological functions that can be expressed in terms of the functional $V^{-1}\int V(\al)$, where $V\equiv \exp[6\al]\,\bv(\al)$. The main of them is $\chi(\al)=d\psi/d\al$.\footnote{The equation for $\bchi\equiv d\al/d\psi\equiv 1/\chi(\al)$ was first published in \cite{ATF}.\,}  Their detailed properties and applications to flat isotropic models are presented in Chapter~5. The conditions for developing and terminating inflationary scenarios are derived, and  some  examples of the inflationary $\chi^2(\al)$ are given. Unlike potential, this function is extremely simple and intuitive: small $\chi^2$ defines inflationary regime; with growing $\chi^2\rightarrow 6$ the ratio~\footnote{The corresponding characteristic function $\rbl(\al)$ is precisely defined in Section~\textbf{4.3}. } of the kinetic energy to the potential $\bv(\al)$ becomes infinite, which actually means the end of inflation.\footnote{Compression of dark matter can be similarly described by considering $\bv(\al)<0, \chi^2>6$.}  This approach somewhat vaguely resembling the "inverse method" of scattering theory is what we mean by speaking of `constructing' cosmologies.

 Finally, a few words on what the reader will not find in this text. In the process of preparing the manuscript all traces of calculations with concrete potentials and cosmological scenarios were eliminated. The aim of such a cutoff is to most clearly present the general \emph{logical structure} of the approach. The reader is invited to test its merits and shortcomings by considering her/his favorite models of `inflating' or `bouncing' universes.

 \section{Cosmological models with scalaron and vecton}
 
 \subsection{On unified theory of gravity, dark energy, and dark matter}

 Here, is given a simple formulation of the minimal affine  extension of general relativity that produces dark energy and dark matter. It was first proposed in \cite{ATF} and further developed in \cite{ATFn} - \cite{ATFg}, where more detailed derivations are described. The `\emph{geometric}' Lagrangian consists of two different parts -- the generalized Ricci curvature $r \equiv r_{ik}\,g^{ik}$ and the Lagrangian derived from the Eddington - Einstein term $\sqrt{\,r}\,$. A generalization of this construction is described in the quoted papers but here I propose to avoid Einstein's ingenious derivation of the non-Riemannian component of the connection with the aid of \emph{ad hoc} variational principle. Today, it looks most appropriate to use the connection that is `\emph{almost Riemannian'}, i.e., having the \emph{geodesics that coincide with Riemannian up to parametrisation}: 
 \be
 \ga^i_{jk}\,\meqdef\,\Ga^i_{jk}(g_{mn}) + \al\,(\de^t_j\,a_k + \de^i_k\,a_j)\,;  \qquad r\,\meqdef\,g^{ik}\,r_{ik} = R - 3\,\al^2 a_i\,a^i \,,
 \label{e0} 
 \ee
 where $\Ga^i_{j\,k}$ is the Christoffel symbol and $R\equiv\,g^{ik}\,R_{ik}$ - the corresponding scalar curvature (note that $g_{00}<0$).
 The fundamental properties of this connection are: 1)~the conformal Weyl tensor  is given by its Riemannian part, 2)~the mass squared of the vecton $a_i$ is positive and arbitrary (as seen in the Lagrangian below).\footnote{In \cite{ATFn}, \cite{ATFs}, I introduced a more general connection by adding one more term, $\beta\,g_{jk}\,a^i$. H.Weyl introduced (1921) the connection with $\beta=\al$, Einstein (1923) considered the case $\beta=-3\al$. The squared masses are negative for both  Weyl and Einstein models because, in (\ref{e0}), $\De\,R = +6\,\al^2 a_i\,a^i\,$ instead of $-3\al^2 a_i\,a^i $.} Note that we omit the terms of the first order in $a_i$ which are proportional to $\nabla^{(g)}_i\,a^i$ and vanish `on the mass shell'.

 The antisymmetric part of the generalized Ricci scalar must be  proportional to $f_{ij}\equiv\p_i\,a_j - \p_j\,a_i\,$, $\,f_{01}=\da_1-a_0^\pr\,$; it requires an additional term in the Lagrangian that must be both `geometric' and `dynamical'. Eddington and later Einstein considered constructing such a Lagrangian using the Ricci curvature $r_{ik}$ or the sum of $s_{ik}$ and $f_{ik}$ that can generate an expression producing the quadratic term $\sim f^2 \equiv f^{ik} f_{ik}$ when $\al^2 a_i\,a^i$  is small. Now one should use dimensional considerations, taking into account that geometric dimensions must be $\textrm{L}^N$ in the system of units $\hbar = c =1$; then $N[a_i] = -1$,  $N[\al] = 0$, $N[f_{ik}] =-2$, etc. Using, with proper reservations, the original ideas of reviews \cite{Ein2} and considerations of more recent papers \cite{ATF}, \cite{ATFn}-\cite{ATFg} we make our final proposal for the \emph{`geo-dynamical'} Lagrangian:
 \be
 2\ka\,{\Lag}_{{\textmd{geo}}}^{(4)}\,=\,\sqrt{-g}\,\bigr\{R - 3\,\al^2 a_i a^i - 2\La\,[\,\det(\de^i_j + \la f^i_j)\,]^\half \,\bigl\}\,,\qquad \kappa\,\meqdef\,8\pi G\,,\quad N[G]=2\,.
 \label{e1}
 \ee
 Here $G$ is Newton's constant, $g=\textrm{det}(g_{ik})$, $\La$ and $\la$ are constants with dimensions $N[\La] = -N[\la]= -2$; parameter $\la$ must be equal to $\sqrt{\ka/\La}$ as will be shown below. The first two terms are produced by the symmetric part of the generalised Ricci tensor $r_{ij}$, its antisymmetric part depends only on the third term.
 In the limit of small $|f_{ij}|$ it is the sum of the `\emph{cosmological constant'} and of the kinetic term of the massive vector field $a_i$, so that
 \be
 2\ka\,{\Lag}_{gAs}^{(4)}\, = \,\sqrt{-g}\,\biggr[R - 2\La - \kappa \biggr(\half F_{ij}F^{ij} + m^2 A_i\,A^i +\, \p_{\,i}\psi\,\p^{\,i}\psi + v(\psi) \biggl) \biggl]\,.
 \label{e2}
 \ee
 The scalar Lagrangian\footnote{Massive scalars appear in dimensional reductions of higher dimensional versions of (\ref{e2}), see \cite{ATFs}-\cite{ATFg} e.a.} is added for comparison (note unusual normalizing of our potential,
 $v\equiv 2V$), $\,\kappa\,m^2 \equiv 3\al^2$; capitals $A_i, F_{ij}$ are used for the standard massive vector field variables.

 For future consideration we supplement the Lagrangian of the \emph{final theory unifying Einstein's gravity, dark matter, and dark energy} with the Lagrangian of the scalaron matter that, at the moment, may be better understood as a standard theory  describing inflation:
  \be
 2\ka\,{\Lag}_{gas}^{(4)}\, = \,  \sqrt{-g}\,\bigr\{R - 2\La\,[\,\det(\de^i_j + \la f^i_j)\,]^\half\,-  \kappa\bigr[m^2\,a_i\,a^i +\, \p_{\,i}\psi\,\p^{\,i}\psi + v(\psi) \bigl] \bigl\}\,.
 \label{eq1}
 \ee
 A large part of this paper is devoted to our exact approach to constructing  scalaron inflation models that in next papers will be applied to studies of the \emph{`geo-dynamical'} model (\ref{e1}).
 
 \subsection{Spherically symmetric reduction}

 We begin with the spherically symmetric theory and briefly summarize its reductions to general (one-dimensional) cosmological models. Our notation for the spherical metric is
 \be
 ds^2 =\,e^{2\alpha}\,dr^2 +\,e^{2\beta}\,d\Omega^2 (\theta , \phi) -
 e^{2\gamma} dt^2 +\,2\,e^{2\delta} dr dt \,,
 \label{eq2}
 \ee
 where  $\alpha, \beta, \gamma, \delta$ depend on $(t, r)$ and  $d\Omega^2$ is the metric on the 2-sphere. Then the two-dimensional reduction of Lagrangian (\ref{eq1}) that is split in three parts can easily be found:\,\footnote{A detailed description of the reduction procedure is given in our earlier work, see \cite{ATF3}. We omit the total derivatives not affecting the equations of motion and \emph{include the factor} $2\kappa$ \emph{into definitions of Lagrangians\,}.}
 \be
 2\kappa\,\cL_g^{(2)} \equiv e^{ 2\beta\,+\,\al\,+\,\gamma}\,[\,e^{-2\al}(2\beta'^2\,+\, 4\beta' \gamma')\,-\,e^{-2\ga}\,(2\dbe^2\,+\, 4\dbe \dal)\,+\,2\bk\,e^{-2\beta}\,]\,,
 \label{eq3}
 \ee
 \be
 2\kappa\,\cL_a^{(2)} = -\,e^{2\beta\,+\,\alpha\,+\,\ga}
  \bigl\{2\,\La\,[1\,-\,\la^2\,e^{-2(\al+\ga)}\, (\da_1-a_0^\pr)^2\,]^{\half}\,
  +\,\kappa\,m^2 (e^{-2\ga}\,a_0^2\,-\,e^{-2\al}\,a_1^2) \bigr\}\,,
 \label{eq5}
 \ee
 \be
 2\kappa\,\cL_s^{(2)}=\,\kappa\,e^{2\beta\,+\,\alpha\,+\,\ga}\,[\,e^{-2\gamma} \,\dpsi^2 - e^{-2\alpha} {\psi^\pr\,}^2\,-\,v(\psi)\,] \,,
 \label{eq4}
 \ee
 where all functions depend on $(t,r)$,  $\,a^2=a_1^2-a_0^2$. The reduction of the vecton part in (\ref{e2}) is
 \be
 2\kappa\,\cL_A^{(2)} = -\,e^{2\beta\,+\,\alpha\,+\,\ga} \bigl\{2\,\La\,-\,\kappa\,[\,e^{-2(\al+\ga)}F_{01}^2\,+\, m^2\,(e^{-2\ga}A_0^2\,-\,e^{-2\al}A_1^2\,)]\bigr\}
 \label{eq6}
 \ee
 Comparing this with the expansion of $\La$-term in (\ref{eq5}) we find (identifying $a_i$ with $A_i$):
 \be
 2\La\,-\,\la^2\La\,e^{-2(\al+\ga)}f_{01}^2\,+\,...= 2\La\,-\,\kappa\,e^{-2(\al+\ga)}F_{01}^2\,\qquad \textrm{and thus} \quad \la^2 = \kappa /\La\,.
 \label{A1}
 \ee
 It follows that the `\emph{geo-dynamical}' Lagrangian depends on two new observable,  dimensional parameters, $\La$ and $m^2$, which naturally characterize dark energy (cosmological constant $\La$ when $f_{ij}\rightarrow 0$), and presence of dark matter (in domains where  $a_i\neq0$).

 Neglecting the $\de$-term may result in losing an important restriction on further dimensional reduction of equations (\ref{eq3})-(\ref{eq5}). In fact, any dependence on $\delta$ can only be  removed (e.g., by taking the formal limit $\delta \rightarrow -\infty$) after the variation in this parameter. This gives the \emph{momentum constraint}, which in this limit is simply \be
 -{\dot{\beta}}^{\prime} - \dot{\beta} {\beta}^{\prime} +
 \dot{\alpha} {\beta}^{\prime} + \dot{\beta} {\gamma}^{\prime} \,\,
  \propto \,T^{(m)}_{01}\,=\,\p\cL^{(m)}/\p g^{01}\,\, \quad  \textrm{for} \quad  g^{01} \rightarrow 0\,\,\,.
 \label{A2}
 \ee
 Here $T^{(m)}_{01}$ is the energy-momentum tensor of scalar matter, $\dot{\psi} {\psi}^{\pr}$, or vector matter, $a_0 a_1$; vector kinetic terms in models (\ref{e2}) and (\ref{eq1}) do not contribute. Eq.(\ref{A2}) is very important in deriving 1-dimensional reductions to cosmologies, static states (black holes), and nonlinear waves of matter coupled to gravity. In fact, Lagrangians (\ref{eq3})-(\ref{eq4}) describe the spherically-symmetric \emph{static} states, nonlinear \emph{waves} of scalar or vector matter, and inhomogeneous anisotropic spherical cosmologies interacting with gravity, as well as very complex processes of their formation in two dimensions. Exact solution of Einstein`s equations produced by (\ref{eq3})-(\ref{A2}) is in general impossible although there exists a class of nontrivial explicitly integrable models with scalar matter, in which  such solutions were derived and studied.\footnote{In \cite{ATF1}-\cite{VDATF2} a class of 1+1 dimensional gravity-scalar models with exponential potentials is analytically solved and explicitly reduced to one-dimensional static, cosmological, and wave solutions.}
 They are beautifully simple, and the only problem is that their beauty does  not belong to real world.

 \subsection{One-dimensional reductions}

 Returning to hopefully real problems, it should be stressed that equation (\ref{A2}) is very simple for models (\ref{eq4}),(\ref{eq5}): $T^{(m)}_{01} \propto (\dpsi \prpsi + \da\,\pra)$, which is zero when $a$, $\psi$ depend on one variable. To find possible homogeneous cosmological models we thus require  $\psi=\psi(t)$, $a \equiv a(t)$, and make further reductions by separating the variables $t$ and $r$ in the metric:\footnote{By a coordinate transformation it is possible to make $\alpha_1(r)=
 \gamma_0(t)=0$ and we omit these functions.}
 \be
 \alpha = \alpha_0(t) + \alpha_1(r) , \quad \beta = \beta_0(t) +
 \beta_1(r) , \quad \gamma = \gamma_0(t) + \gamma_1(r) \,.
 \label{A3}
 \ee
 This separation is also applicable to finding static solutions, for which we require , $\psi=\psi(r)$, $a_0 = a(r)$. The momentum constraint severely restricts possible cosmological (static) models that can be derived from the two-dimensional Lagrangian. Additional restrictions may also emerge from the equations of motion, up to demanding $v_\psi\equiv0$ or even $v\equiv0$.
 
 Inserting (\ref{A3}) in Eq.(\ref{A2}) with  $T^{(m)}_{01}=0$ and ignoring the irrelevant solution $\dot{\beta}= {\beta}^{\prime}=0$ we find that all momentum constraint restrictions are given by simple equation
 \be
 \,\,\, {\gamma_1}^{\prime} / {\beta_1}^{\prime} \, +
 \dot{\alpha_0} / \dot{\beta_0} \, = 1\,.
 \label{A0}
 \ee
 In the cosmological case, the main possible solutions are: $\beta^\prime = \gamma^\prime=0$ (`\emph{general anisotropic' cosmologies}) and $\dot{\alpha}=\dot{\beta},\, \gamma^\prime=0$ that gives the isotropic FLRW-cosmology. We call the `\emph{special' anisotropic} cosmologies those with $\beta^\prime = \gamma^\prime= \dal=0$, see a `toy model' in Section~\textbf{7.2}. The 1-dimensional theories reduced by conditions $\psi^\prime=0$, (\ref{A2}), (\ref{A3}) describe  homogeneous cosmologies with scalaron $\psi(t)$ if we their three-dimensional curvature is constant
 \be
 R^{(3)} \equiv 2\bk\,e^{-2 \beta_1} - 2\,(\,2{\beta_1}'' +\, 3 {\beta_1'}^2)\,\equiv\, -6k\,.
 \label{A4}
 \ee
 The three-dimensional subspace is isotropic if
 \be
 R^1_1=-2(\,{\beta_1}'' +\, {\beta_1'}^2)\,=\,R^2_2=\bk\,e^{-2 \beta_1} -({\beta_1}'' +\,2{\beta_1'}^2)=R^3_3\,.
 \label{A5}
 \ee
 If $\prbe \neq 0$, the two conditions are obviously equivalent to one,
 \be
 {\beta_1'}^2 - \bk \,e^{-2 \beta_1} \,=\,k \,,
 \label{A6}
 \ee
 When $\prbe \equiv 0$, the cosmology is spatially homogeneous as $R^{(3)} = 2\bk$, while the spatial isotropy requires that it must be flat, i.e. $\bk=0$. It follows that the homogeneity (constant curvature) follows from the isotropy condition (\ref{A5}).

 The homogeneity and isotropy also directly follow from the conditions $\dot{\alpha}=\dot{\beta},\, \gamma^\prime=0$ and from the requirement of separation in the equations of motion. These matters were discussed in some detail in papers quoted above, where it is shown that the `general anisotropic' cosmology (with $\beta^\prime = \gamma^\prime=0$) can be spatially isotropic if $\bk=0$. Then also $k=0$ but nevertheless, for the general solution, $\dal \neq \dbe$ as is demonstrated below. In this case, there exists the unique solution with $\dot{\alpha}=\dot{\beta}$. The 4-dimensional solution then necessary satisfies the condition $\gamma^\prime=0$  and therefore it is the true FLRW-type solution.\footnote{Note that usually the term `isotropic' also implies that $\dot{\alpha}=\dot{\beta}$, see, e.g., \cite{Dau}.}

 From now on we ignore the subtleties related to higher-dimensional interpretation of the solutions and only consider (one-dimensional) cosmological dynamical equations. The aim is to demonstrate that under fairly general conditions the cosmological equations have solutions describing expanding universes which become asymptotically flat and isotropic. The immediate goal is to construct inflationary models. In the scalar model, a class of `inflationary potentials' is explicitly constructed by use of exact general solutions in $\al$-formulation (Section~5.3)). We believe a similar approach can be applied to models (\ref{e2}) and (\ref{eq1}).

 \section{General equations of very early cosmology}
 
 \subsection{Equations describing scalaron and vecton cosmologies}

 Let us first introduce cosmological reductions of theories (\ref{eq1}). As was shown in \cite{ATFn}, the reduced 1-dimensional  cosmological Lagrangian can be written as the sum of three reduced  Lagrangians -- gravitational, $2\kappa\,\cL_g^{(2)} \Rightarrow L_g$, plus similarly defined $L_v\,$, $L_s\,$:
 \be
 L_c = L_g\,+\,L_s\,+\,L_v\,,\quad \textrm{where} \quad L_g\,= -e^{2\beta\,+\,\al - \gamma} (2\dbe^2 + 4\dbe \dal) - 6k\,e^{\al+\ga}\,.
 \label{eq10}
 \ee
 The standard Lagrangians for massive vector and scalar fields ($A\equiv A_1(t),\,\psi\equiv \psi(t)$) are:
 \be
 L_v\,=\,e^{2\beta + \alpha +\gamma}\,\{-2\La\,\,+\, \kappa\, e^{-\,2\al} \bigl[\,e^{-2\gamma}\dA^2 - m^2 A^2\bigr]\}\,,\quad L_s\,=\, \kappa\,e^{2\beta\,+\,\alpha+\ga} \bigl[e^{-2\gamma} \dpsi^2\,-\,v(\psi)\bigr]\,,
 \label{eq10a}
 \ee
 where the $\La$-term in the vector theory appears quite naturally in the context of our "remake" of Einstein's model, if we apply the small field approximation. In the scalar case we tacitly assume that there may exist a constant part $2\La$ of the potential, $v = 2\La + v_1(\psi)$. The vecton partial Lagrangian of the \emph{`affine cosmology'} is more convenient to use if derived by reduction $\cL_a^{(2)} \Rightarrow  L_a$ with replacing $2\La/2\kappa$ by $\la^{-2}$ and denoting $a\equiv a_1(t)$:
 \be
 L_a\,\equiv\,-e^{2\beta\,+\,\alpha\,+\,\ga}  \bigl[\,\la^{-2}\,\sqrt{1\,-\,\la^2\,\da^2\,e^{-2(\al+\ga)}}\,+\, \half\,m^2 a^2 e^{-2\al} \bigr]\,.
 \label{eq10b}
 \ee
 Recall that for small $\da^2$ the Lagrangian can be approximated by $L_v\,$. As $L_a$ cannot be directly applied to present-day cosmology one cannot immediately estimate $\La$ by identifying it with the `measured' cosmological constant.\footnote{To have a `normal' field $A(t)$ we must suppose that $\La>0$, $m^2>0$.} However, when the kinetic energy of dark matter becomes small  enough it presumably may be related to observable estimates of dark energy. The second new fundamental constant $m^2$ cannot be theoretically related to $\La$ or other known mass parameters. To simplify treatment of the nonlinearity effects  in the vecton Lagrangian we propose a nonlinear Lagrangian like (\ref{eq10b}), linearizing of which gives $L_s$:
 \be
 L_{sa}\,\equiv\,-\,e^{2\beta\,+\,\alpha\,+\,\ga}
 \bigl[\la^{-2}\,\sqrt{1\,-\,\la^2\,\dvphi^2\,e^{-2\ga}}\,+\,\half\,m^2 \f^2 \bigr]\,,
 \label{eq10c}
 \ee
 where the scalar field $\f$ replaces the above $\psi$. 
 Although this model does not correctly represent anisotropy of the vecton model it is as useful as the linear scalaron cosmologies, and similar theories were considered in connection with brane  cosmologies.\footnote{See, e.g., \cite{Silver1}-\cite{Cop}, where such models are attributed to DBI. In fact, a model like (\ref{eq10b}) first appeared (with wrong interpretation) in Einstein's paper \cite{Einstein1}, which was influenced by Eddington's idea.}

 In summary, we plan to consider four separate models: gravity with `linear' scalaron (the standard scalaron cosmology), gravity with linear vecton (vecton cosmology), gravity with nonlinear vecton (affine cosmology), and somewhat artificial nonlinear scalaron cosmology. Our program is to solve these theories in full generality. This is obviously impossible and we first simplify them by neglecting anisotropy and curvature, $k$, having in mind that for large enough Universe these become small and probably could be estimated by some perturbation theory. In the standard approach, these effects are neglected from the very beginning. We show why this is justified in case of inflationary solutions for linear scalaron and, hopefully, for vecton models (\ref{eq10a}) and estimate the small anisotropic and curvature corrections to exact isotropic solutions. In fact, in this paper we partially solve the scalar one  and show that anisotropy and curvature effects are asymptotically small (compare also to extremely anisotropic solvable model in Section~\textbf{7.2}). However, nonlinear models  are much more tricky and undoubtedly  require a significantly different treatment that can only be hinted here.

 We begin with writing the equations of motion for the complete Lagrangian $L_c$. To present them in a form convenient for considering the anisotropic case, we introduce notation
 \be
 3\rho\,\meqdef\,(\alpha + 2\beta)\,,\quad 3\sigma\,\meqdef\,
 (\beta-\alpha)\,,\quad \al\,\meqdef\,\rho-2\sig\,;\quad 3\,A_{\pm}\,\meqdef\, e^{-2\al}\,(\dA^2 \pm m^2 e^{2\gamma} A^2)\,.
 \label{eq11}
 \ee
 Then the exact Lagrangian for the vecton-plus-scalaron cosmology given by equations (\ref{eq10})-({\ref{eq10a}) can be equivalently rewritten as ($\La$-term is included in $v(\psi)$,
 \be
 L_c = e^{3\rho - \gamma} (- 6\drho^2\,+\,6\dsig^2\,+\,\dpsi^2 + e^{-2\al}\dA^2) \,-\,e^{3\rho + \gamma}\,[\,v(\psi)\,+\, m^2\,e^{-2\al}A^2] \,-\,6\,k\,e^{\,\rho-2\sig+\ga}\,.
 \label{eq12}
 \ee
 Here $e^{-\gamma}$ is the Lagrange multiplier, variations in $(-\ga)$ yield the \emph{energy constraint}:
 \be
 H_c \equiv e^{3\rho - \gamma}(- 6\drho^2 + 6\dsig^2 + \dpsi^2)\,+\,e^{\rho+4\sig-\ga}(\dA^2 + m^2 e^{2\ga} A^2)\,+\, e^{3\rho + \gamma}\,v(\psi)\,+\,6\,k\,e^{\,\rho - 2\sigma+\ga}=0. 
 \label{eq13}
 \ee
 It is equivalent to vanishing of the total Hamiltonian to be written below. As in any gauge theory with one constraint of this type, we can choose one gauge fixing condition (the most obvious ones are $\gamma = 0$, $\gamma = \alpha$, $\gamma = 3\rho$). The other dynamical equations are:
 \be
 \ddot{\rho}\,+ (3\,\drho - \dga)\,\drho - e^{2\gamma}\,v(\psi)/2 \,=\,2k\,e^{2\ga-2(\rho + \sigma)} \,+\,(3 A_+\,-\,A_-)/4 \,,
 \label{eq15}
 \ee
 \be
 \ddot{\psi}\,+\,(3\,\drho\,-\,\dga)\,\dpsi\,+\,e^{2 \gamma}\, v^\pr(\psi)/2 \,=\,0\,,
 \label{eq14a}
 \ee
 \be
 \quad \ddot{\sigma}\,+\,(3\,\drho\,-\,\dga)\,\dsig\,=\,k\,e^{2\ga-2(\rho + \sigma)}\,+\,A_{-} \,,
  \label{eq16}
 \ee
 \be
 \ddA\,+\,(\drho\,+\,4\dsig\,-\,\dga)\,\dA\,+\, m^2A\,e^{2\gamma}\,=\,0 \,.
  \label{eq14b}
 \ee

 Taking account of the constraint (\ref{eq13}) one can replace Eq.(\ref{eq15}) by equation
 \be
 \ddot{\rho}\,-\,\drho \dga\,+\,3\,\dsig^2\,+\,\dpsi^2/2 \,=\,-k\,e^{2\ga - 2(\rho+\sigma)}\,-\,(3 A_+\,+\,A_-)/4\,
  \label{eq15a}
 \ee
 that is independent of the potential $v(\psi)$. Note that the sum $\,3A_+ \pm A_- \geq 0\,$ if $\,m^2>0$, and can vanish only if $\dA = A = 0$. Then   Eq.(\ref{eq15a}) gives the important general inequality for the `Hubble  function' $\drho(t)$.\footnote{Actually, $h(\tau) \equiv\drho(t)$ is not the standard Hubble "parameter", $H(t)$, defined for flat homogeneous cosmologies. In other words, one should better identify $H(t)$ with our $\dal(t)$. Instead, we mostly use the generalized Hubble function defined as  $h(\rho)=\xi(\rho)\equiv \drho(t)$ (see Eq.(\ref{mom4}) below).} Introducing the `gauge invariant time' parameter $\tau$ by $e^{\ga}d/dt \equiv d/d\tau\,$, Eq.(\ref{eq15a}) can be rewritten in a gauge-invariant form and gives the inequality
 \be
 dh(\tau)/d\tau = -[\,3\,\dsig^2\,+\,\dpsi^2/2\,+ \,k\,e^{-2(\rho+\sigma)} \,+\,(3 A_{+} \,+\,A_{-})/4\,]\,<0\,,\quad \textrm{if}\quad\,k\geq\,0\,,
 \label{eq15b}
 \ee   
 where the dots temporally denote $d/d\tau$. This equation defines strong restrictions on the behavior of the Hubble parameter that are independent of the scalar potential (including the cosmological constant) but significantly depend on all kinetic energies and on the potential energy of the vecton. If $h(\tau)$ is asymptotically constant, all terms in the r.h.s. of equation  (\ref{eq15a}) \emph{must vanish} for $\tau \rightarrow +\infty\,$. This shows that $\dh(\tau)$ asymptotically decreases if and only if $\dsig^2, \dpsi^2, A_+, A_-$ asymptotically vanish ($\dA^2, A^2$ are exponentially bounded above).\footnote{If $\al(t) \rightarrow +\infty$, it is a reasonable assumption but a study of asymptotic behavior of $\al, \dA, A$ is needed.}

 From this discussion it must be clear that asymptotically constant Hubble function ($3\drho^2$ in (\ref{eq13})) requires a constant term $v_0$ in the potential. As will be shown below,  the construction of scalar inflationary models suggest approximately constant potentials in the inflationary domain and thus introduction of $2\La$ is only necessary in the approximate Lagrangian (\ref{e2}). Strictly speaking, the discussed equations may be applicable on a narrow interval, from Planck-order time to times near the end of inflation, when new degrees of freedom become important.
 Nevertheless, as will be shown by considering exact solutions, it makes sense to study asymptotic behavior of the solutions up to the point where one of the characteristic functions of the inflation begins fast growing. It may possibly be interpreted as a transition from inflation to scenarios in which many new degrees of freedom are ignited and the real universe starts emerging. Ideas on possible description of the universe before and at the end of inflation are not discussed here. We mainly study the problems of solving the dynamical equations and the problem of constructing the only unknown element -- potential.\footnote{The case of affine cosmology (\ref{eq10b}) is discussed separately and only briefly.}
  
 \subsection{The importance of being canonical}

 Let us begin with summarizing general properties of Eqs.(\ref{eq13})-(\ref{eq15a}). Obviously the simplest are the matter equations (\ref{eq14a})-(\ref{eq14b}) that play important roles in all approaches to solving our dynamical systems. The pure scalar isotropic cosmology, obtained by taking $A\equiv 0$, $k=0$, and $\sig\equiv 0$, was exactly solved in the $\al$-\emph{representation} for any gauge choice, \cite{ATFnew}-\cite{ATFgen}. The formal exact solution with $k\neq 0,\,\sig\equiv0$ is shown (in this paper) to be an attractor in the asymptotic region $\al \rightarrow \infty$. In fact, we here construct an asymptotic iterative approximation to this \emph{attractor}, which shows that the most general inflationary solutions are only asymptotically flat and isotropic. This possibly indicates that more complex essentially anisotropic cosmologies  aught to be carefully studied in spite of their non-integrability.\footnote{The classic investigation on stability of cosmological states is \cite{LifKh}. Cosmologies considered here apparently require a special study. Presumably, the flat isotropic cosmologies are attractors.}

 To better understand general structure of equations (\ref{eq13})-(\ref{eq15a}) it is useful to have a look at the canonical momenta, 
 \be
 (p_\rho,\,p_\psi,\,p_\sig) = 2\,e^{3\rho-\ga}(-6\drho,\,\dpsi, \,6\,\dsig)\,,\quad p_A = 2\,e^{\rho + 4\sig -\ga} \dA\,,
 \label{mom1}
 \ee
 and to rewrite the Hamiltonian (\ref{eq13}) into the canonical form: 
 \be
 e^{-\ga}H_c^{\textrm{can}} = \frac{e^{-3\rho}}{24}(- p_\rho^2 + p_\sig^2 + 6p_\psi^2  + 6p_A^2\,e^{2(\rho-2\sig)})\,+\,v(\psi)\,e^{3\rho} \,+\, 6k\,e^{\rho-2\sig} \,+\, m^2 A^2\,e^{\rho+4\sig}\,.
 \label{mom2}
 \ee
 In this way, equations (\ref{eq15})-(\ref{eq16}) with $A\equiv 0$ and $k=0$ were simplified and solved in \cite{ATFnew}, \cite{ATFgen}.
 
 Treating nonlinear vecton theory (\ref{eq10b}) looks much more difficult and requires a separate investigation. Here, we only sketch its Hamiltonian formulation.
 Defining
 \be
 p_a \equiv dL_a/d{\da} = e^{2\beta}\,\da\,(e^{\al+\ga}-\la^2 \da^2)^{-1/2}, \qquad H_a \equiv p_a \da - L_a(p_a,a)\,,
 \label{mom3}
 \ee
 it is not difficult to derive exact effective Hamiltonian $H_a$ (recall that $\la^{-2}=\La/\kappa$):
 \be
 H_a = e^{\al+\ga} [\,\la^{-1}\,\sqrt{p_a^2 + M^2_{\textrm{eff}}}\,+\,\half\,m^2 a^2 e^{\,2(\beta -\al)}\,], \qquad M_{\textrm{eff}}\equiv \frac{e^{2\beta}}{\la}\,.
 \label{mom3a}
 \ee
 A similar but simpler Hamiltonian can be derived in nonlinear scalar case (\ref{eq10c}):
 \be
 H_{sa} = e^{\ga} [\,\la^{-1}\,\sqrt{p_{\f}^2 + m^2_{\textrm{eff}}}\,+\,e^{2\beta+\al}\,m^2 \f^2/2\,]\,, \quad p_{\f}= e^{2\beta+\al}\,\dvphi\,(e^{2\ga}-\la^2 \dvphi^2)^{- 1/2}\,,
 \label{mom3b}
 \ee
 where $m_{\textrm{eff}}\equiv e^{3\al}/\la\,$. When the gravitational field is constant, both models can be explicitly integrated. In fact, in this approximation they are rather similar. Significant simplifications may appear in the isotropic approximation $\beta=\al$ though the scalar model is essentially isotropic. The most important common property of both models is that the energy of the matter fields ($a$ or $\f$) may become infinite when their `velocities' ($\da$ or $\df$) are finite. In the gauge $\ga+\al=0$, the limiting value of the maximal `velocity' is $\da_{max} \equiv \sqrt{\kappa/\La}$. Therefore, the dynamics of the field $a(t)$ looks like that of a 1-dimensional relativistic particle in an external field. This was observed in \cite{ATFg} and may become a crucial property of our nonlinear models, which will possibly turn out to be a kind of \textit{cosmic `accelerator' of dark matter}.

 In \cite{ATFnew}, we discussed only the simplest gauge choices, which in the present general model look like $\ga + c\rho =0$.  The  gauge $\ga+\al=0$, that simplifies the vecton Hamiltonian, was not used in cosmology but it is the textbook gauge for the Schwarzschild black hole. When $A\equiv\,0,\,k=0$, the gauge $\ga-3\rho = 0$ is most useful in the linear scalaron cosmology (unlike the "standard" gauge $\ga=0$). More general gauges are $\,g(p,q;\ga)=0$.\footnote{Here $q =\psi, \rho, \sig,  A$ and $p$ -- the corresponding momenta. As $q$ are independent of $\ga$, a simple and safe gauge condition is $\,g\,(q;\ga)=0\,$ with $\,\p_\ga g\neq 0$. The restriction on the general $g$ is $\,\p_p\,g\,\p_\ga p\,+\,\p_\ga g \neq 0$.} Thus if we try to choose a gauge in which r.h.s. of (\ref{eq16}) vanishes we must express $\dA^2$ (in  $A_-$) in terms of $p_A\,$. Then we find that this condition does not fix $\ga$ because all the terms are proportional to $e^{2\ga}$. In cosmological and static models the `standard' gauge $\ga^\pr=0$ and the `light cone' gauge are most popular but here we mostly use the gauge independent solutions first introduced in  \cite{ATFnew} (linear isotropic scalaron models) and in \cite{ATFgen} (general linear scalaron).

 \bigskip
 \section{Gauge independent solution of scalaron cosmology}
 
 \subsection{Definitions and solutions}

 Let us rewrite $\psi$-equations (\ref{eq12})-(\ref{eq16}), with $A\equiv0$, in the first-order form using, instead of $\rho,\,\psi,\,\sig$, the momentum-like variables $\xi\,,\eta\,,\zeta$ treated as functions of $\,\rho$\,($d/dt \equiv \xi\,d/d\rho$):\,\footnote{In \cite{ATFnew}, we have exactly solved equations for $\xi(\al),\,\eta(\al)$ and approximately -- the corresponding equations for $\xi(\psi),\,\eta(\psi)$. The `transition' function  $\chi(\al)\equiv\psi^\pr(\al)$ connects the $\psi$ and $\al$ pictures. In  discussing the general anisotropic cosmologies it is more appropriate to replace $\chi(\al)$ by $\chi(\rho)\equiv\psi^\pr(\rho)$. }
 \be
 (\drho,\,\dpsi,\,\dsig)\,\meqdef\, [\,\xi(\rho)\,,\eta(\rho)\,,\zeta(\rho)\,] = [\,\xi(\rho),\,\xi\,\psi^\pr(\rho),\,\xi\,\sig^\pr(\rho)\,]\,\meqdef\,  \xi(\rho)[\,1,\,\chi (\rho),\,\om(\rho)]\,.
 \label{mom4}
 \ee
 Here $\chi(\rho)\,\meqdef\,\eta/\xi = \psi^\pr(\rho)$ and $\om(\rho)\,\meqdef\,\zeta/\xi =\sig^\pr(\rho)$ are gauge independent functions, the integrals of which, $\psi(\rho)$ and $\sig(\rho)$, give a \emph{portrait} of the scalaron cosmology that defines its characteristics. In paper \cite{ATFnew} we thoroughly studied $\chi(\al)$ and its another incarnation $\bchi(\psi) \equiv d\xi/d\psi = 1/\chi[\,\rho(\psi)]$. Here we derive and study solely the dynamical function defined in (\ref{mom4}). When $\sig \equiv 0$ the fundamental function $\chi(\rho)$ can be derived if we take as the input either potential $\bv(\rho)$ or the \emph{generalized Hubble function} $H(\rho) \equiv \xi(\rho)$.\,\footnote{When $k\,\neq 0$, equation (\ref{eq16}) has no solution with $\dsig \equiv 0$, but $\dsig$ may be exponentially small if $\rho \rightarrow +\infty$. For this reason, flat isotropic models with $\sig \equiv 0$ and $k=0$ are generally meaningful as explained above.}
 
 Transforming the dynamical equations into first-order form requires a slight generalization of \cite{ATFnew}. The scalaron dynamical system consists of equations (\ref{eq15})-(\ref{eq16}).
 Constraint (\ref{eq13}), being its integral, is automatically satisfied when the arbitrary integration constant is chosen zero. Then the only problem remaining is the presence of $v^\pr(\psi)$ and of the $\sig$-dependence when $k\neq0$. The formal substitution $\bv(\rho) = v[\psi(\rho)])$ allows us to solve all equations in the $\rho$-version and, as soon as $\chi(\rho)$ ist derived, it will be possible to reconstruct the potential in the $\psi$-version:  $v(\psi)=\bv[\rho(\psi)]$ for arbitrary $\bv(\rho)$. The obvious relation,
 \be
 v^\pr(\psi) = \frac{dv}{d\psi} = \frac{dv}{d\rho} \frac{d\rho}{d\psi} = \bv^\pr(\rho)\frac{\xi}{\eta} = \bv^\pr(\rho) /\chi(\rho)\,,
 \label{mom5}
 \ee
 then solves the $v(\psi)$ problem. Comparing the definitions (\ref{mom4})-(\ref{mom5}) and their discussion with the corresponding consideration in \cite{ATFnew}, it is not difficult to find that (\ref{eq15})-(\ref{eq14a}) (with $A_\pm=0$, $k=0$) are linear differential equations for $\xi^2,\,\eta^2,\,\zeta^2$ in any gauge $\ga(\rho)$ and can be exactly solved for any potential $\bv(\rho)$. The $k$-terms are proportional to $\exp{[-2\sig(\rho)]}$ but, in cosmological consideration, we may suppose that $\sig \ll 1$ as will be argued below.

 Rewriting the dynamical equations in terms of the \emph{positive gauge-invariant functions},
 \be
 \textbf{S}(\rho)\equiv[\,x(\rho),\,y(\rho),\,z(\rho)\,]\,\meqdef\, \exp{(6\rho-2\ga)}\, [\,\xi^2(\rho)\,,\eta^2(\rho),\,\zeta^2(\rho)\,]\,,
 \label{mom6}
 \ee
 we see that (\ref{eq14a}) gives the only equation independent on $\sig$ and $k$-term: 
 \be
 y^\pr(\rho)+V^\pr(\rho) - 6V(\rho) = 0\,, \qquad V\equiv e^{6\rho}\,\bv(\rho)\,.
 \label{mom7}
 \ee
 Depending only on the fundamental potential $V(\rho)$, the solutions of this equation form the basis of the most general solution $\textbf{S}(\rho)$. A study of the properties of this general solution will give important restrictions on the potential $\bv(\rho)$. Two other equations depend on the curvature term and on $\sig$, and so the whole system is obviously not closed:
 \be
 x^\pr(\rho) - V(\rho) = 4k\,e^{4\rho -2\sig},\,\,\,\, \textrm{(a)} \qquad z^\pr(\rho) = 2k\,e^{4\rho -2\sig}\,\sig^\pr(\rho).\,\,\,\, \textrm{(b)}
 \label{mom8}
 \ee
 The additional equation that makes it closed is the \emph{definition}: $z/x\,\equiv\,\zeta^2/\xi^2\,\equiv\,\om^2\,\equiv\,{\sig^\pr}^2$. This will allow us to show that $\sig$ satisfies a closed equation, the solution of which can be estimated and derived perturbatively in the limit of small anisotropy at large $\rho$.

 We first find the formal solution of equations (\ref{mom7})-(\ref{mom8})  plus constraint (\ref{eq13}), which is
 \be
 6\,x(\rho)\,=\,y(\rho) + V(\rho)\,+\,6\,z(\rho)\,+\,6k\,e^{4\rho-2\sig}\,.
  \label{mom9}
 \ee
 To this end, it is sufficient to integrate equations (\ref{mom7}), (\ref{mom8}\,a) and use (\ref{mom9}) ($\,i=x,y\,$):
 \be
 y = 6I(\rho) - V(\rho),\,\,\,x = I(\rho)\,+\,i_x(\rho); \quad I \equiv C_y\,+\int V(\rho),\,\,\,i_x = C_z + 4k\int e^{4\rho-2\sig},
 \label{mom10}
 \ee
 \be
 z \equiv x(\rho)\,{\sig^\pr}^2(\rho) \,=\, i_x(\rho)\,-\,k\,e^{4\rho-2\sig(\rho)}\,\equiv\,C_z \,+\,2k\int \sig^\pr(\rho)\,e^{4\rho-2\sig(\rho)}\,,
 \label{mom11}
 \ee
 where $C_z\equiv C_x-C_y\,$ and integration in all the integrals is taken over the interval [$\rho_0,\,\rho\,$], with  $\rho_0 \geq -\infty\,$.\footnote{Here and below we assume that $\rho_0=-\infty\,$. It is possible due to exponential vanishing of the integrands.} Eq.(\ref{mom11}) for $\sig(\rho)$ will be discussed below. Supposing that $\sig \ll 1$ one can derive $\sig$ by some sort of iterations. The first approximation for $x$, obtained for $\sig=0$, coincides with the result of \cite{ATFnew}, because $z\equiv 0$ and therefore $C_x=C_y$. The new function, which is a gauge-invariant measure of anisotropy, is given by $\om^2 = z(\rho)/x(\rho)=(\sig^\pr)^2$. From (\ref{mom11}) we see that, when $k=0$, there exists the solution $\om^2(\rho)=C_z/x$, and thus, if  $V(\rho)>0$ is fast growing with $\rho$, there may exist non-isotropic solutions that become isotropic for large $\rho$. For the true isotropic solutions $\sig\equiv 0$ and $C_x=C_y$, but one can keep $k\neq 0$ because Eq.(\ref{mom11}) is trivially satisfied and $k\neq 0$ corrections are exponentially small. The flat isotropic solutions ($k=\sig^\pr \equiv 0$) will be studied below in great detail. The general solution given by Eqs.(\ref{mom9})-(\ref{mom11}) may become isotropic only asymptotically, when $\rho\rightarrow +\infty$.  The isotropic solutions are apparently the limiting or enveloping ones for solutions with $C_z\neq 0$, $k\neq 0$.

 In \emph{summary}, there may exist four types of solutions:  1)~when $k=\sig^\pr=C_z=0$, the isotropic solution formally coincides with the standard FLRW cosmology;  2)~$k=0,\,\sig^\pr\neq 0,\, C_z\neq 0$ gives the simplest anisotropic solution, which for a wide class of the potentials $\bv(\rho)$ becomes isotropic at large $\rho$, because  $\sig^\pr \rightarrow 0$ for $\rho \rightarrow +\infty$;  3)~$k\neq 0,\,\sig^\pr = 0,\, C_z= 0$ is a special isotropic solution; 4)~$k\neq 0,\,\sig^\pr \neq 0$ is the general anisotropic solution; under certain restrictions on $\bv$, the anisotropy $\sig(\rho)$ can vanish at $\rho \rightarrow +\infty$. When $k=0$, these solutions asymptotically coincide with the standard FLRW cosmology.
 
 \subsection{Positivity conditions}

 To complete the solution one should add to equations (\ref{mom7}) - (\ref{mom9}) some \emph{conditions for positivity} of $x(\rho), y(\rho),  z(\rho)$. Not pretending to be most general we give conditions that are most suitable in considering inflationary solutions when $k=0, \sig \equiv 0$. Suppose that $\bv(\rho)>0$ and consider the integral with an arbitrary parameter $\la>0$, $-\infty \leq \rho_0 <\rho:\,$\footnote{This parameter $\la$ depends on the considered models as well as on their dimensional reductions. In Appendix we consider the model with $\la = 4$. Note that in the next equation we use the partial integration that is extremely important for establishing a connection of our approach to the standard one.}
 \be
 y(\rho) \equiv \la\int_{\rho_0}^\rho\,e^{\la \rho}\,\bv(\rho) \,-\, e^{\la\,\rho}\,\bv(\rho) + e^{\la\,\rho_0}\bv(\rho_0)\,.
 \label{mom11a}
 \ee
 When $\la =6 $ and $V=e^{6\rho}\,v(\rho)> 0$ this defines the solution of (\ref{mom7}) with a special choice of arbitrary constant $C_y$ in $(\ref{mom10})$ giving $y(\rho_0)=0$. Its \emph{positivity requires} $\bv^\pr\,(\rho)<0$ because
 \be
 y(\rho)\,=\,-\,\int_{\rho_0}^\rho\,e^{\la\rho}\,\bv^\pr(\rho) \,,\quad y^\pr(\rho)\,=\,-e^{\la \rho}\, \bv^\pr(\rho)>0\,.
 \label{mom11b}
 \ee
 These relations showing positivity both for $y(\rho)$ and $y^\pr(\rho)$ become especially simple and useful in the limit $\rho_0 \rightarrow -\infty$ and will be used in flat isotropic models. Then general positive solution $y(\rho)$, Eq.(\ref{mom10}), is obtained by taking in (\ref{mom11b}) $\la=6$ and adding arbitrary positive constant. The positivity of $x(\rho)$ is similarly secured if $k\geq 0$ (when $k<0$ it may require a large enough values of $\rho >0$). Finally, positivity of $z(\rho)$ is ensured if $k\sig^\pr\geq0$ and $C_z\geq 0$.\footnote{Note that only the scalaron momentum squared, $y(\rho)$, can be safely considered in the interval $-\infty <\rho <0$, because the curvature and anisotropy corrections may become dominant for large negative $\rho$ unless the potential $\bv(\rho)$ has compensating exponentially large terms $\sim e^{-2\rho}$. Anyway, in this domain the classical picture must (possibly) be replaced by a quantum one, which is not discussed in this paper.}

 It should be stressed that our `\emph{general positive}' solution may become negative when $\rho_0$ is finite and that the variety of solutions is immense for general potentials $\bv(\al)$. For instance, \emph{contracting evolution is possible only with negative potentials}. The positivity of such solution is naturally provided by choosing $\rho < \rho_0$ and/or large positive $C_x,\,C_y$. However, $\rho_0$ must be a not very large positive number because of the exponential growth of $V(\rho)$. Anyway, here we mostly discuss solutions describing inflationary-like expansion but the \emph{mathematical formalism is applicable to treating arbitrary potentials and diverse cosmological scenarios}.

 A more general discussion of the positivity conditions was attempted in \cite{ATFgen}. However, the solutions having `negativity domains', with branch points at the ends (in the complex plane $\textbf{C}_{\al}\,$), are not of immediate physics interest at the moment. Existence of such solutions possibly hints at a real necessity of analytic continuation of the solutions as well as of the potential $\bv(\al)$. But this does not look a very promising idea.
 It might be a bit more interesting to consider models in which solutions in the domain $\rho < \rho_0$ are negative. Then one may try to speculate that in this domain they are quantum or at least semiclassical. Attempts to relate classical initial conditions to some quantum models have a long history (see, e.g., \cite{Hal}-\cite{CAF}, e.a.), and recently this interest has been revived (\cite{Linde-new}-\cite{Rubakov3}), e.a.). We stay here on classical grounds and, moreover, formally extend the classical picture to $\al_0 =-\infty$.
 
 \subsection{Three characteristic cosmological functions}
 
 In the general solutions, the curvature and isotropy corrections vanish at $\rho \rightarrow +\infty$. It follows that for inflationary solutions one can neglect the dependence on $k$ and $\om(\rho)$ when $e^{2\rho}$ is large enough. In fact, the anisotropy corrections are smaller than the curvature ones\,\footnote{For $\rho \rightarrow +\infty$, $\om(\rho) \sim e^{-2\rho}$. Thus,  anisotropy corrections, being  multiplied by $ke^{-2\rho}$, must be $\sim e^{-4\rho}$.} and it is meaningful to study the solutions with $\om \equiv 0$, $k\,\neq 0$ for a better understanding the transition to completely isotropic regime. In this \emph{isotropic approximation} it is easy to derive the $\chi$-function, which in general depends also on $\sig(\rho)$. From equations (\ref{mom8})-(\ref{mom11}) we find the following exact and approximate expressions for $\chi^2(\al)$ (always $V>0,\,I>0$):
 \be
 \chi^2\,\meqdef\,\frac{y}{x}\,\equiv\, \frac{6I(\rho)-V(\rho)}{I(\rho)+i_x(\rho)}\, \stackrel{\sig\rightarrow 0}{=}\,\,6\,-\frac{V(\al)\,+\, 6ke^{4\al}}{I(\al)\,+\,ke^{4\al}}\,\,\stackrel{k\rightarrow 0}{=}\,\, 6\,-\frac{V(\al)}{I(\al) }\,,
 \label{mom11c}
 \ee 
 where we always use $\al$ instead of $\rho$ when $\sig\,\equiv\,0$. If, in addition, $k=0$, the expression for $\chi^2$ is equivalent to  constraint (\ref{mom8}), as $x=I(\al)$.  By differentiating Eq.(\ref{mom11c}) in $\al$ and  expressing $I(\al)$, $I^\pr(\al)$ in terms of $V$ and $\hchi^2$ it is easy to find \emph{the main equation} for $\chi^2(\al)$ when $\om=0$: 
 \be
 [\chi^2(\al)]^{\,\pr}\,[1+6\,k\,e^{-2\al}/\bv\,] = (\chi^2-6)\,[\,\chi^2(1+ 4\,k\,e^{-2\al}/\bv)+\,\bl^{\,\pr}(\al)]\,.
 \label{mom11d}
 \ee
 Here we introduce the \emph{important function} $\bl^{\,\pr}(\al)\equiv \bv^{\,\pr}(\al)/\bv(\al)$ that completely defines the behaviour of the fundamental function $\chi^2(\al)$ in the `classical' domain, i.e., for large enough positive values of $2\al$ (and $2\rho$). This means that solutions of physics interest and, in particular, the inflationary solutions are `\emph{asymptotically invariant}' with respect to the scale transformations of the cosmological potential. By the way, up to now we used the naturally normalized transition function $\chi(\rho)$  for which $\chi^2<6$. It will be often more convenient to normalize it to unity and introduce its \emph{hatted version} $\hchi(\rho)\,$: $\,\hchi^2(\rho)\,\equiv\,\chi^2/6$.

 Next Section mostly deals with even more important characteristic cosmological function, the \emph{gauge independent ratio of the scalaron kinetic to potential energy}:
 \be
 \rbl(\rho)\,\meqdef\,\frac{y(\rho)}{V(\rho)}\equiv\, \frac{6I(\rho)}{V(\rho)}-1\,\equiv\,\frac{\hchi^2(\rho)\, [1+\kbl(\rho)]}{1-\hchi^2(\rho)-\om^2(\rho)}\,;\qquad \kbl(\rho)\,\meqdef\,\frac{6\,k}{\bv(\rho)}\,e^{-2(\rho+\sig)}\,,
 \label{mom11e}
 \ee 
 which is \emph{independent} of $\om$ and $\kbl$.\footnote{The apparent dependence of r.h.s on these function simply compensates the real dependence of $\hchi^2$ on $\kbl$.}
 In addition, we introduce another important function $\wbl(\al)$ that coincides with $\chi^2(\al)/6$ when one takes $\om\equiv 0,\,\kbl=0$ in Eq.(\ref{mom11e}). Then, obviously:
 \be
 \wbl(\rho)=\frac{\rbl(\rho)}{\rbl+1}\,,\quad \rbl(\rho)=\frac{\wbl(\rho)}{1-\wbl}\,, \quad
 \wbl(\rho)\,\meqdef\,1-\frac{V(\rho)}{6I(\rho)}\,=\,1 - \frac{1}{6}\, [\ln\,I(\rho)]^\pr\,,
 \label{mom11ee}
 \ee
 where all the functions are independent of $\om$, $\kbl$,   positive, and $\wbl(\rho)<1$. Actually, $\wbl$ is equivalent to $\rbl$ and (\ref{mom11e}) gives exact expressions for $\hchi^2$ in terms of either functions.
 Relation (\ref{mom11ee}) \emph{is crucial for constructing inflationary models} providing for positive $[\wbl\,(\al)]^{\,\pr}$ a \emph{"graceful exit"} from small-$\hchi$-inflationary regime when $\wbl\,(\al)$ is approaching unity. Indeed, $\,\rbl\simeq \wbl\,$ for small $\wbl$, and  $\rbl\,\rightarrow \infty$, if  $\,\wbl \rightarrow 1$. Condition $\rbl\ll 1$ is the most general \emph{criterion for inflationary regime} and $\rbl(\rho)$ satisfies the \emph{exact equation} independent on $\sig(\rho)$ and $\kbl$,
 \be
 \rbl^\pr(\rho) \,+\,[6 + \prbl(\rho)]\,\rbl(\rho)= -\bl^\pr(\rho)\,, \qquad \rbl^\pr \,=-[6\,\rbl+(\rbl+1)\,\bl^\pr]\,. 
 \label{mom11dd}
 \ee
 It directly follows from definition (\ref{mom11e}) and is equivalent to (\ref{mom11d}) when $\kbl=\sig \equiv 0\,$, so that\footnote{When $\kbl=0$, it is sufficient to suppose that $C_z=0$, see (\ref{mom10}-\ref{mom11}).}
 \be
 \wbl^\pr(\al) = -[1-\wbl(\al)]\,[6\wbl(\al) + \bl^\pr(\al)]\,\meqdef\,[1-\wbl(\al)]\,\Theta(\al).
 \label{mom11dw}
 \ee
 By definition, this equation is valid for all $\kbl, \om$ but does not coincide with (\ref{mom11d}). The functions $\rbl, \wbl$ are fundamental and most important for formulating main scenarios of very early cosmologies. Finding approximate expression for $\om(\rho)$ would allow to reconstruct the general expression for  $\hchi^2$ by inverting Eq.(\ref{mom11e}) and to estimate small anisotropic and curvature terms, which in future could possibly be used to find small observable effects.
 
 Sometimes it may be convenient to use the analogous but more complicated function,
 \be
 \rbl_x(\rho)\,\meqdef\,6\,x(\rho)\,V^{-1} = [\rbl\,(\rho) + 1 + \kbl(\rho)]\,[1-\om^2(\rho)]^{-1}\,\,,
 \label{mom11f}
 \ee
 which is somewhat simpler than $\hchi^2(\rho)\equiv \rbl/\rbl_x$ but does not give new information. In applications to inflationary scenarios we mostly use characteristic functions $\rbl(\rho)$ and $\wbl\equiv \rbl\,(\rbl+1)^{-1}$. 
 Note that only the results obtained by using $\rbl(\rho)$ are exact for all $\om$ and $\kbl$.

 Let us summarize our \emph{most general mathematical description of the scalaron cosmology}. If $\sig\equiv 0$, it is sufficient to know just one function $V(\al)$ or, equivalently, $I(\al)$. Then one can derive explicit expressions for all the physically important cosmological functions by using equations (\ref{mom9})-(\ref{mom10}), (\ref{mom11c}), (\ref{mom11e})-(\ref{mom11f}). The same can be done once we know one of the three functions: $\hchi^2(\al)$, $\rbl(\al)$, or $\rbl_x(\al)$. To reconstruct all information on cosmology in the anisotropic case one must derive the `anisotropy' function $\om(\rho)$ by solving equation (\ref{mom11}), which is obviously equivalent to the nonlinear equation derived by its differentiating:
 \be
 \sig^{\pp}(\rho)+(\ln\sqrt{x})^\pr\,\sig^\pr(\rho)\,=\, k\,[x(\rho)]^{-1}\,e^{4\rho-2\sig}\,.
 \label{mom11h}
 \ee
 When $k\neq 0$, the gauge invariant generalized `Hubble function' $\sqrt{x(\rho)}$ implicitly depends on $\sig(\rho)$ and the exact solution cannot be derived. The first iteration is discussed in Appendix~\textbf{7.1}. For $k\neq 0$ the anisotropic correction $i_x(\rho)$ can be derived by simple iterations starting with $\sig\equiv 0$. When $k=0$ but $C_z\neq 0$, there exists a special anisotropic solution with asymptotic behaviour $\sig\,\sim\,e^{-3\rho}$ that will not be discussed here.\footnote{It should be stressed that the solution with $\sig\equiv 0$, $k\neq0$ is a consistent special  solution (not the general). }
 
 \bigskip
 \section{On cosmological scenarios and potential}
 
 \subsection{Main properties of $\rbl(\al)$, $\wbl(\al)$,  and  $\Theta(\al)$} 

 As was argued in \cite{ATFnew}, in the present `$\al$-approach' the most fundamental entity is the function $\chi^2(\al)$. When anisotropy vanishes, $\sig(\al)\equiv 0$, the function $\chi(\al)$ completely define the portrait $\psi(\al)$ that can be derived with one integration. Moreover, once this function is known, one can in principle derive the complete solution of the cosmological equations. This is obvious when $k=0$ but can be generalized to $k\neq 0$, if we find the general $\chi(\rho)$.
 
 In physics interpretation of the model parameters, the most important functions are $\chi(\rho)$, $\rbl(\rho)$, and a peculiar function $\Theta(\al)\equiv -(\chi^2(\al)+\bl^{\,\pr}(\al))$.\footnote{On using variables $\rho$ and $\al$: in general  we write $\rbl(\rho)$, $\wbl(\rho)$ if the case $k, \om \neq 0$ is discussed. We write $\rbl(\al)$ and $\wbl(\al)$ in the context of FRLW cosmology. When possible, the dependence is omitted.} It first appeared in the equation for $\chi^2(\al)$ derived in \cite{ATFnew} and is the $k=0$ reduction of Eq.(\ref{mom11d}). The equation for $\wbl(\al)$ is:
 \be
 \wbl^\pr(\al)\,=\,[1-\wbl]\,\Theta(\rho)\,\equiv (\rbl+1)^{-1}\,\Theta(\rho)\,.
 \label{mom11i}
 \ee
 It is not difficult to find the general, independent of $k$ and $\sig$ relations (see (\ref{mom11e}-\ref{mom11dd})):
 \be
 \Theta(\rho)\,=\,\ln\,(I/ V)^{\,\pr}\,=\, [\,\ln(\rbl+1)]^{\,\pr}\,, \quad \textrm{i.e.} \quad \rbl^\pr(\rho) \equiv (\rbl+1)\,\Theta(\rho)\,,
 \label{mom11j}
 \ee
 where $\rho$ is recovered because (\ref{mom11dd}) and both these relations are independent of $\om, k$. As $\,0<\wbl(\al)<1$, the first derivatives of functions $\wbl(\al)$ and $\rbl(\al)$ have evidently the same sign as $\Theta(\al)$, and this fact is of importance for describing inflationary models. The  inflationary expansion is realised when $\rbl^\pr\,,\wbl^\pr>0$ while, e.g.,  $6\wbl(\al)\lesssim 1$ and
 $6\rbl(\al)\lesssim 1$. If  $\wbl(\al)$ monotonically grows with $\al$, the  $\rbl(\al)$ function grows up to $+\infty$ when $\wbl(\al)$ approaches its upper bound $+1$, as can be seen from Eq.(\ref{mom11ee}). Also recall that $\bl^\pr<0$ \emph{is necessary for} $\Theta >0$.

 The $\Theta(\rho)$ function satisfies the evident first-order differential equation,
 \be
 -\Theta^\pr(\rho) = \Theta^2 + L^\pr(\rho)\,\Theta + \bl^{\pp}(\rho) \equiv (\Theta - \Theta_+)(\Theta - \Theta_-)\,,
 \label{mom11k}
 \ee
 where $L^\pr(\rho) \equiv [\,\ln V(\rho)\,]^\pr = 6+\bl^\pr(\rho)>0$.\,\footnote{We usually assume that $L^\pr(\rho)>0$, the conditions for positivity are discussed later.} The roots ($\Theta_+$, $\Theta_-$), depending on $\prbl(\rho)$ and  $\bl^{\pp}(\rho)$ are both negative if $\bl^{\pp}>0$ and may have opposite signs when $\bl^{\pp}<0$. In the last case, $\Theta_+ >0$  and $\Theta^\pr$ is positive when $\Theta < \Theta_+ =-\bl^\pr/L^\pr+...\,$.  This consideration can be applied  to easily finding the sign  of $\wbl^{\pr\pr}$ in terms of $\Theta,\,\prbl,\,\bl^{\pp}$.  Using equations (\ref{mom11i}),   (\ref{mom11k}) we get
 \be
 \wbl^{\pp} = (1-\wbl)\,(\Theta^\pr\,-\,\Theta^2) =
  -(1-\wbl)\,[2\Theta^2 + L^\pr\,\Theta + \bl^{\pp}]\,,
 \label{mom11m}
 \ee
 find the roots, and factorize.  The statement on the signs of the roots $\Theta_+$, $\Theta_-$ is equally applicable to this equation thus defining the sign of $\wbl^{\pp}$. Denoting its roots $\tThe_{\pm}(\rho)$, we find
 \be
 4\,\tThe_{\pm} = \,=\, \pm\,L^\pr(\sqrt{1-8\,\bl^{\pp}/(L^\pr)^2}\mp 1), \quad  \tThe_{+} = -4\bl^{\pp}/L^\pr(1+...) \geq 0\,,\,\,\tThe_{-}<0\,,
 \label{mom12}
 \ee
 if $\,L^\pr(\rho) \equiv (6+\bl^\pr)>0\,$ and $\,\bl^{\pp}(\rho)\leq 0$. It is easy to derive the signs of $\Theta$ for different intervals of $\rho$. It follows that $\wbl^{\pp}$ is positive for $\Theta < \tThe_+$ and changes the sign when $\Theta =\tThe_+\,$.

 Presumably, in the inflationary domain,  $\wbl^{\pp}$ is positive but later it may become negative. After a \emph{zero} of $\wbl^{\pp}$ (at the point of \emph{inflection}, $\al_i$) it becomes possible that at a point $\rho_m >\rho_i$ both $\wbl^{\pr}$ and $\Theta$ change their  sign. Alternatively,  $\Theta(\rho)$ will remain positive, and then $\wbl(\rho)\rightarrow 1$ while $\rbl(\rho) \rightarrow +\infty$. Such a singularity (corresponding to a "big bang"?) may look physically unacceptable but, in fact, a similar scenario may occur when the maximum value $\wbl(\rho_m) \equiv \wbl_m$ defines a large enough maximum of the matter energy function, $\rbl(\rho_m)= \wbl_m/(1-\wbl_m)$.

 \subsection{On constructing inflationary characteristic functions}

 The restrictions on $\rbl(\rho)\,,\wbl(\rho)$ and on derived from them $\Theta,\,\wbl^\pr,\,\wbl^{\pp}$, \emph{encode crucial properties of the main cosmological scenarios and, therefore, of the potential} $V(\al)$.     
 All these functions depend on one function, $6L_I(\rho)\equiv \ln I(\rho)$ and its derivatives, $L^\pr_I =V/6I$, etc. In particular:
 \be
 \rbl(\rho)+1 = (L^\pr_I)^{-1}>1, \quad 0<(1-\wbl) = L^\pr_I(\rho)<1, \quad \wbl^\pr = -L^{\pp}_I(\rho),
 \label{mom19}
 \ee
 \be
 \Theta = -L^{\pp}_I\,(L^\pr_I)^{-1},  \quad \rbl^\pr=-L^{\pp}_I \,(L^\pr_I)^{-2}\,,\qquad \wbl^{\pp}=-L^{(3)}_I(\rho),
 \label{mom20}
 \ee
 where we use Eqs.(\ref{mom11ee}, \ref{mom11i}, \ref{mom11j}) and, of course, $I^\pr(\rho)=V>0$. Now it follows that
 \be
 \rbl^\pr, \wbl^\pr, \Theta \gtreqqless 0\,\,\Longleftrightarrow\, L^{\pp}_I \lesseqqgtr 0, \qquad  \wbl^{\pp}  \gtrless 0 \Longleftrightarrow L_I^{(3)}\lessgtr 0,
 \label{mom20a}
 \ee
 and that $\,\wbl^{\pp}(\rho_i)=0$ is equivalent to $L^{(3)}_I(\rho_i)=0$.\,\footnote{The point $\rho_i$ in which $\wbl^{\pp}=0$ but $\wbl^\pr\neq 0$ is called the \emph{inflection} point. Its coordinates are $\{\,\rho_i,\,w_i\}$.}
 It is also easy to derive the expressions
 \be
 \Theta^\pr = [\wbl^{\pp}(1-\wbl)+(\wbl^\pr)^2](1-\wbl)^{-2}, \qquad
 \rbl^{\pp}=[\wbl^{\pp}(1-\wbl)+2(\wbl^\pr)^2](1-\wbl)^{-3}\,,
 \label{mom20b}
 \ee
 that well describe the behaviour of $\Theta$ and $\rbl$ when $\wbl$ approaches its maximum. In the inflationary domain  $\wbl^{\pp},\,\rbl^{\pp},\,\Theta^\pr$ must be positive before the inflection point. Though it is not extremely important, note that the inflection points of $\wbl$ and $\rbl$ are significantly   different.

 The next paragraph shows how these simple relations allow one to `draw' the potential corresponding to different scenarios by guessing appropriate characteristic functions. In the inflationary interval and beyond it, up to the first zero of $\Theta(\al)$, the functions $\rbl$ and $\wbl$ monotonically grow to the common maximum at $\al_\textrm{max}$.  The crucial reason for emerging inflation is the \emph{smallness} of $\rbl$ (and $\wbl$) in a \emph{large enough $\al$-interval} . The crucial property for the exit from inflationary regime is a \emph{fast enough growth} of $\rbl$ in some interval of $\al$, so that in the end $\wbl$ \emph{becomes close to the upper bound} $\wbl = 1$. A description of the graphic construction is given in the next paragraph as well as in Section~\textbf{7.1}.\footnote{\,One may also consider a class of functions $L_I$ satisfying inequalities (\ref{mom19}) and restrictions implied by any selected scenario. This will allow analytic realizations of different possible scenarios by constructing $L_I(\al)$.}

 Near the end of inflationary domain both $\wbl(\al)$ and $\rbl(\al)$ begin to grow. For example, take as the end-point $\al_{out}\,$, in which $\wbl=1/6\,$ 
 In the interval  $(\al_{\textrm{ex}}, \al_{\textrm{max}})$, $\wbl(\al)$ monotonically increases, while $\rbl(\al)$  grows very fast if $\,(1-\wbl_\textrm{max})\,$  is small. The picture for $\al>\al_{\textrm{max}}$ is not controlled by the model as it ignores emergent new physics phenomena. However, mathematical consistency requires $\rbl(\al)$ to be small for $\al>\al_{\,\textrm{fin}}\,$, see \textbf{5.3}. The parameters, (which are essentially conventional as well as their descriptive definitions `small', `large'), 
 \bdm
 -\infty\qquad\al_{\textrm{in}}\,\longrightarrow\,\al_{\,\textrm{out}} \,\longrightarrow\al_{\textrm{ex}}\longrightarrow\,\, \al_m\equiv\al_{\textrm{max}} \longrightarrow\,\, \al_{\,\textrm{fin}}\,\dashrightarrow\,\qquad +\infty
 \edm
 \bdm
 \,\,0\qquad \quad\wbl_{\textrm{in}}\longrightarrow\,\wbl_\textrm{out} \longrightarrow\wbl_\textrm{ex}\longrightarrow\wbl_\textrm{m}\equiv \wbl_\textrm{max} \longrightarrow \wbl_{\,\textrm{fin}}\, \dashrightarrow\qquad\quad 0,
 \edm
 can be visualized following prescriptions of \textbf{7.1}, which are \emph{not quite realistic}\,\footnote{Let us agree that "in" means \emph{initial}, "out" means \emph{beginning of the end of inflation}, "ex" - \emph{exit from inflation}, "fin" - \emph{a physical limit of applicability of the model} that may remain mathematically consistent.}, especially, because the curves need not be symmetric w.r.t. vertical line $<\al_{\textrm{max}}\,, \wbl_\textrm{max}>$ and the length of interval  $(\al_\textrm{out}\,,\,\al_\textrm{max})$ need not be of order one (the curves for $\al>\al_\textrm{max}$ are optional).

 To have the minimal number of the parameters let us suppose that the curve $\wbl(\al)$ has only one inflection point, $\al_i$, in the \emph{post-inflationary} interval $(\al_{\,\textrm{out}}, \al_{\textrm{max}})$\,\footnote{The behavior of $\wbl$ on $(\al_{\,\textrm{in}}, \al_{\textrm{out}})$ is defined by expansion of \textbf{5.4} which cannot account for zeroes of $\wbl^{\pp}$.} and therefore $\wbl^{\pp}(\al)\gtrless 0$ when $\al\lessgtr \al_i\,$. The coordinates of this point, $\{\al_i,\wbl_i\}$, define the tangent line, which gives important information on the shape of the post-inflationary curve $\wbl(\al)$. After the inflection point there are two possible shapes. \textbf{1})$\,\wbl^\pr(\al_m)=0$ for some $\al_m = \al_{\textrm{max}}>\al_i\,$, in which $L^{\pp}_I=0$ but $\wbl^{\pp}=-L^{(3)}_I<0$. To construct the complete curve one needs to know $\wbl^\pr(\al_{\,\textrm{out}})$, to choose a point $\{\al_i,\wbl_i\}$ that lies above the tangent line at $\{\al_{\,\textrm{out}},\wbl_{\,\textrm{out}}\}$, and to secure that $\wbl_i^\pr\,>\,\wbl^\pr(\al_{\,\textrm{out}})>0$. Finally, if we take $\wbl^{\pp}_m=0$; then according to (\ref{mom20a}-\ref{mom20b}):
 \be
 \rbl^\pr_m = \wbl^\pr_m = \Theta_m = \rbl^{\pp}_m = \wbl^{\pp}_m = \Theta^\pr_m = 0, \qquad \al=\al_m\,.
 \label{mom20c}
 \ee

 It can be seen that this `graphic' approach is a significant complement to the pure algebraic one based on Eqs.(\ref{mom19}-\ref{mom20b}). It allows one to discuss a \emph{`soft exit'} from inflation. The resulting remarkably simple construction is based on minimum of fairly natural assumptions realizing our rigorous definition of what is inflation. In Section~\textbf{5.4} we briefly present a perturbative approach that allows to establish contacts with the standard approximation and with few existing observational data. One may hope that the proposed mathematical formulation for possible mechanisms of post-inflationary physics may be used in attempts to find some new observable effects. Nore that the curves defined on the interval $(\al_{\textrm{max}},\, \al_{\,\textrm{fin}})$ may prove to be quite different from those discussed above and in Section~\textbf{7.1}. Even the length of the interval itself is not known and at the moment cannot be even estimated. 

 The simplest possible scenario is the following: \textbf{2})$\,\wbl^\pr(\al)$ and $\rbl^\pr(\al)$ remain positive for all $\al>\al_i\,$, while $\wbl(\al)\rightarrow 1$, when $\,\al\rightarrow +\infty$, and $\wbl(\al)\rightarrow 0$, when $\,\al\rightarrow -\infty$. This case may be called a `\emph{hard exit}' from inflation. The simplest realization of it is given by $\wbl_0$:
 \be
 (a): \,\wbl_0=\frac{1}{\pi}\arctan(\la\bal)+\half\,, \quad (b): \,\wbl_1=\theta(\bal_m-\bal)\,\wbl_0(\bal) + \theta(\bal-\bal_m)\,\wbl_0(2\bal_m-\bal)\,.
 \label{mom20d}
 \ee
 Here $\,\theta(\al)$ is a regularized step-function, $\bal=\al-\al_i\,$, so that $\wbl_0^{\pp}(0)=0.$ The first function, $\wbl_0$, looks  un-physical and is not further discussed  here.\footnote{In Section~\textbf{5.3}, we argue that its behavior at $\al \rightarrow  +\infty$ disagrees with our restrictions on potential $\bv(\al)$.} The second one, $\wbl_1$, coincides with $\wbl_0$ in the limit $\al_m \rightarrow +\infty$. For finite $\al_m\,$, it has a bell-shaped form with discontinuity of $\wbl_1^\pr(\bal)$ at $\bal_m\equiv \al_m-\al_i$. By smoothing it out we get an acceptable curve, $\wbl_2(\bal)$, having $\wbl_2(\bal_m)$ close to the upper limit $1$ and essentially depending on large enough parameter $\la$.\,\footnote{With large $\la$ the effective width of the `bell' can be made small while $\wbl_{max}$ close to the maximal. Then the corresponding figure of $\rbl$ will be very high and sharp, so forming a sketch of `Eiffel Tower' (ET-sketch).}

 Up to the point $\al_i\,$, this picture is fairly simple and does not require any extension of the considered model, except the presumably `quantum' interval $-\infty<\al< \al_{\,\textrm{classical}}\,$. In what follows we usually extend the lower limits of $\al$-integrations to $-\infty$ using the fact that the integrals have the `damping' factor $\exp(6\al)$. Much more problematic is the interval $\al>\al_m\,$, where simple models are strictly speaking inapplicable. Although it is not difficult to find potentials providing inflationary scenarios with apparently possible post-inflationary exit, realizing these scenarios requires tools beyond the present elementary considerations. 
 
 \bigskip
 \subsection{Potentials from characteristic functions}

 The simplest global construction of the potentials defining a fairly general cosmological scenarios is the following. The general $\hchi^2(\rho)$ with $\om\equiv0$ follows from Eqs.(\ref{mom10}), (\ref{mom11c}):
 \be
 \hchi^2 = [6I(\rho)/V(\rho)-1]\,[\,6I/V + \ka(\rho)]^{-1},\qquad \ka(\rho)\, \equiv 6\,i_x(\rho)\, V^{-1}(\rho)\,.
 \label{mom21}
 \ee
 Expressing $V/I$ in terms of $\hchi^2$ and remembering that $V=I^\pr(\rho)$ it is easy to find that
 \be
 \bv(\rho) = \bv_0\,\frac{1-\hchi^2}{1+\ka\,\hchi^2}\, \exp\,6\biggl[-\rho+\int\frac{1-\hchi^2(\rho)}{1+\ka\,\hchi^2}\biggr]\,.
  \label{mom22}
 \ee
 In the flat isotropic limit $\ka(\rho)\equiv 0$, one can replace $\hchi^2(\rho)$ by $\wbl(\al)$ and get a $\ka$-independent exact expression of $\bv(\rho)$ in terms of $\rbl(\rho)$ valid for the general anisotropic model. As a matter of fact, $\bv(\rho)$ can be directly derived from exact relations $\rbl(\rho) =6I/V-1$ or Eq.(\ref{mom11dd}),
 \be
 \bv(\rho) = \bv_0 (1+\rbl(\rho))^{-1}\,
 \exp \int\,\frac{-6\,\rbl(\rho)}{1+\rbl(\rho)}\,\equiv\,\bv_0 (1+\rbl(\rho))^{-1}\,\exp [-6\int\wbl(\rho)]\,.
 \label{mom23}
 \ee
 which is exponentially decreasing when $\rbl \rightarrow +\infty$ ($\wbl \rightarrow 1$). This behavior should be \emph{rejected} according to general restrictions on asymptotic behaviour of $\bv(\rho)$. In other words, one must consider a soft exit, in which both $\rbl(\rho)$ and $\wbl(\al)$ reach a maximum and start fast falling.\footnote{Remember that at large $\rho$ the difference between $\rho$ and $\al$ can be neglected.}  This restriction on potentials agrees with our intuition,  prohibiting infinities in the physically meaningful functions, and makes inflationary scenarios both more complex and more interesting. Anyway, after the exit from inflation, the whole model becomes dubious.

 The potential can simply be derived from (\ref{mom11dd}) or (\ref{mom11j}), which can be rewritten as
 \be
 [\,\ln V(\rho)]^\pr \equiv (6+\bl^\pr) = (6-\rbl^\pr) (\rbl+1)^{-1}\,,\qquad \bl^\pr=-[(\ln\rbl)^\pr+6]\,\wbl(\rho)\,.
 \label{mom24}
 \ee
 This gives the simple but important \emph{theorem}: inequalities $\rbl^\pr(\rho)\leq 6$ and $6+\bl^\pr(\rho)\geq 0$ are equivalent. The statement is rather nontrivial if you remember that $\rbl^\pr(\rho)>0$ is equivalent to $\Theta >0$, which requires $(-\bl^\pr) >6\wbl >0$,  and so $\bl^\pr$ must stay in the  interval $(6\wbl, 6)$. From (\ref{mom24}) it also follows that $\bl^\pr <0$, if and only if $\,\rbl^\pr>-6\rbl\,$. This is the \emph{strongest restriction} on the choice of possible potentials $\bv(\rho)$. The sufficient conditions for existence of inflation are: $\rbl(\rho) \ll 1$ on a large enough interval $(\rho_1, \rho_2)$ and $\,0<\rbl^\pr \leq 6\,\rbl\,$. As is demonstrated by the above construction of the potential, these conditions are also necessary.

 The second important restriction on the cosmological models comes from the Hubble function $h(\rho)$. As seen from equations (\ref{eq15a}-\ref{eq15b}) it may become constant only asymptotically, when $\tau, \rho \rightarrow +\infty$. Exact solution (\ref{mom10}) giving $\xi^2 \equiv x(\rho) e^{2\ga - 6\rho}$ in terms of the integral of the potential is less convenient for estimating global properties of Hubble`s function, and it is better to express $h(\rho)$ by using Eqs.(\ref{mom11f}), (\ref{mom24}). With $\kbl\,=\,\om^2=0$ and standard gauge $\ga=0$:
 \be
 6\,h^2(\al) = \bv(\al)\,(\rbl(\al) + 1) = \bv_0\,\exp[-6\int\wbl(\al)]\,;\qquad \bv(\al) = 6\,h^2(\al) + \frac{dh^2}{d\al}\,.
 \label{mom25}
 \ee 
 Like the potential, our generalized Hubble function is approximately constant in the inflationary domain where $\wbl(\al)\ll 1$.  Unlike (\ref{mom23}-\ref{mom24}), Eq.(\ref{mom25}) is not exact and should be corrected when $\al$ is not large enough to neglect the curvature and anisotropy corrections.

 The corrections can easily be derived from Eq.(\ref{mom11f}). We consider the large interval of $\al$ in which these corrections can be neglected but $\wbl(\al)$ is very small. In this interval the Hubble function is approximately constant and $\,(\ln h^2)^\pr = -6\wbl$ vanishes with $\wbl$.  This is rather a strong restriction on possible behavior of $\wbl$ in the post inflationary interval. The simplest realisation is our `Eiffel-Tower' construction discussed in \textbf{5.2}. Probably, the formulas in this subsection must be significantly changed near $\,\al_{\,\textrm{fin}}$ or earlier. This subject is not  discussed here because, in our opinion, \emph{the exit from inflation and further events should better be treated in the frame of the vecton theory} of Section~\textbf{2.1}. Nevertheless, arguments on smallness of $\wbl$ and $\rbl$, based on expressions (\ref{mom23}, \ref{mom25}), support the discussed above `ET-like scenarios.

 In the flat isotropic model, the expressions for $\bv(\al)$ and $h^2(\al)$ in terms of $\hchi^2(\al) = \wbl$ were first derived in \cite{ATFnew} although the approximate flatness of $\bv(\al)$ was not discussed. It follows from them that $\bv $ and $h^2$ are approximately constant in the inflationary domain, where $\hchi^2$ and $\rbl$ are small. In addition to the present-day \emph{observational evidence in favor of `nearly-flat' potential} (see e.g.\cite{Martin18}, \cite{Martin}-\cite{Martine}) it is worth stressing that in this domain $\rbl^\pr(\rho)>0$, from which it follows $\bv^\pr(\rho)<0$ and thus $\bv \neq \bv_0$ on any interval. Possibly, it may fall dawn to not very small minimum $\bv_{min}>0$ and then starts rising when $\rbl^\pr \leq -6\rbl$.

 \subsection{Perturbative expansion in flat, isotropic models}
 
 Finally, a few remarks on the perturbative expansion for $\rbl$ introduced in \cite{ATFnew}, Sec.~4.3. By carrying out obvious partial integrations in the definition (\ref{mom11e}),
 \be
 \Sig_0(\rho) \equiv 1+\rbl(\rho) \equiv \frac{6e^{-6\rho}}{\bv(\rho)}
 {\int_{-\infty}^{\rho}e^{6\rho}\,\bv(\rho) \,=\,\frac{e^{-6\rho}}{\bv(\rho)}}\bigg[e^{6\rho}\,\bv - \int_{-\infty}^{\rho}e^{6\rho}\,\bv^\pr(\rho)\bigg],
 \label{mom12}
 \ee
 one can derive the important expression, which also defines the main scale-invariant characteristic function of any cosmology, $\Sig_1(\rho)$:
 \be
 \rbl(\rho)\,\equiv\, \Sigma_1(\rho) = \sum_1^{\infty} (-1)^n\, \frac{\bv^{(n)}(\rho)}{6^n\,\bv(\rho)}\,\equiv \,\sum_1^{\infty} \ve^n\,E_n(\rho)\,,\quad \ve \equiv -1/6 \,.
 \label{mom13}
  \ee
 Here $E_n(\rho)$ can be expressed in terms of polynomials of the logarithmic derivatives of the potential, $\,\bl^{(m)}(\rho)\equiv d^m/dr^m \,[\,\ln \bv(\rho)]\,$ by the simple recurrence relation,
 \be
 E_{n+1}(\rho) = E_n^\pr(\rho) + \bl^\pr(\rho)\,E_n(\rho),\,\quad E_1 = \bl^\pr\,, \quad E_2\,=\,(\prbl)^2 + \bl^{\pp}\,,\quad E_3 = \bl^{(3)} +3 l^{\pp}\bl^\pr +(\bl^\pr)^3.
 \label{mom13a}
 \ee
 It is also convenient to use for $n=0$ notation $E_0\,$, $\,\Sig_0 \equiv 1+\Sig_1\,$, and $n=0,1,2,...\,$,
 \be
 \Sig_n (\rho) \equiv \sum_n^\infty \ve^m\,E_m, \quad \Sig_{n+1} \,=\,\Sig_n - \ve^n\,E_n; \qquad -\Sig^\pr_n (\rho) \equiv (6+\bl^\pr)\Sig_n + \ve^{n-1}\,E_n\,.
 \label{mom13b}
 \ee
 Expansion (\ref{mom13}) generalizes equation (78) of \cite{ATFnew}, the basic tool for constructing \emph{inflationary perturbation theory} and for relating our approach to the standard one. It is easy to write the expansions  of $\rbl(\rho), \wbl(\rho), \Theta(\rho)$. For example, expanding $\rbl(\rho)$ in Eq.(\ref{mom11dd}) one finds:
 \be
 \rbl^\pr(\rho) = \Sig^{\,\pr}_1(\rho)\,=\,\ve\,\bl^{\pp}(\rho)\,+\, \ve^2\, [\,\bl^{(3)}(\rho)\,+\,2\bl^\pr(\rho)\,\bl^{\pp}(\rho)] \,+\, \Sig_3^{\,\pr}(\rho)\,,
 \label{mom13c}
 \ee
 which can be extended to any order of $\ve$. The expansion of $\wbl(\rho)$ simply follows from Eq.(\ref{mom11ee}). The expansion for $\Theta(\rho)$ can be derived from Eq.(\ref{mom11j}). A useful simple expression,
 \be
 (1+\Sig_1)^{-1}\,\equiv \sum_0^{N-1}\,(-1)^n \,\Sig_1^n\,+\, (-\Sig_1)^N\,(1+\Sig_1)^{-1}\,,
 \label{mom13d}
 \ee
 will allow to derive all these perturbative expansions in powers of $\ve$. For example, to derive the expansion of $\wbl$ up to the terms of order $\leq \ve^N\,$, it is sufficient to express all functions $\Sig_n$ in terms $E_n$ and keep all relevant $E_m$ in expansion of all terms in expansion (\ref{mom13d}). Thus,
 \be
 \wbl\,\rbl^{-1} = (1+\rbl)^{-1} = 1-\ve E_1 + \ve^2 (E_1^2 - E_2) + \textrm{O}(\ve^3) = 1\,-\,\ve \bl^\pr +\ve^2 [2\,(\bl^\pr)^2 - \bl^{\pp}] + \textrm{O}(\ve^3)\,,
 \label{mom13e}
 \ee
 and multiplying this by $\ve(E_1+\ve E_2 +\ve^2 E_3)$ we find the first terms of expansion of $\wbl$, which also allow to immediately derive the expansion of $\,\Theta \equiv - (6\wbl+\bl^\pr)=-\bl^{\pp}/6 + ...\,$.

 Convergence of the series (\ref{mom13}) strongly depends on $\bv(\rho)$. We may expect that $\bv(\rho)$ is a decreasing function of $\rho$. But, even the simplest potential $(\rho + \rho_0)^{-1}$ defines non-convergent but only asymptotic series (\ref{mom13}).\footnote{In short, $\,E_n(\rho) \rightarrow 0\,$ for fixed $n$ and $\,\rho\rightarrow +\infty\,$, but it grows with $n$. Although the potential is supposed indefinitely differentiable, only a finite number of its terms may be taken into account for any fixed $\rho$.} For the exponential potential $e^{-\la \rho}$ the series is convergent for $\la <6$, but exponential potentials $\bv(\rho)$ are not acceptable in our present $\al$-approach.\footnote{Concentrating on exponential potentials and on exponential expansions of cosmological solutions was a serious deficiency of paper \cite{ATF} that did not allow me, at that time, to discover the $\al$ approach.}

 \textbf{\emph{Criteria for inflation on a long interval.}} $(\rho_0, \rho_1)$:~\textbf{1}.\,First and most important condition is $0<\rbl \ll 1$ (equivalent to $0<\wbl \ll 1$) and $0<\rbl^\pr(\rho) \ll 1$ (equivalent to $0<\wbl^\pr \ll 1$). \textbf{2}.~From these conditions and Eq.(\ref{mom11dd}) it follows that $\bl^\pr<0$ and $|\bl^\pr|\ll 1$. \textbf{3}.~Equations (\ref{mom11i}-\ref{mom11j}) show that $\Theta$ is small. As can be seen from Eq.(\ref{mom11m}), $\Theta^\pr\ll 1$ is equivalent to $\wbl^{\pp}\ll 1$. \textbf{4}.~The last condition is rather natural to impose on $\wbl$ but not on $\rbl$ which can grow indefinitely if $\wbl$ is close to its upper limit.
 In other words, in the inflationary interval all the three functions are supposed to be small and smooth. This means that the corresponding potential $\bv(\rho)$ is slowly and smoothly diminishing. In particular, Eq.(\ref{mom13c}) demonstrates that $\bl^{\pp}<0$ (see also (\ref{mom11k}-\ref{mom11m}).

 \textbf{\emph{Perturbative inflationary `parameters'.}} Our general perturbation theory is especially effective in the inflationary domain, as presented in detail in \cite{ATFnew}, where the reader may find definitions and perturbative expressions for the standard inflationary parameters. The first terms of the characteristic functions are: $\rbl(\rho) = -\bl^\pr(\rho)/6\,+...$; $\,\Theta(\rho) = -\bl^{\pp}(\rho)/6 \,+...\,$. As the higher derivatives are supposed in this interval much smaller, these `parameters' may be considered approximately constant. The small (almost constant) parameters of the standard approach, which are  essentially $[\ln v(\psi)]^\pr,\, [\ln v(\psi)]^{\pp}$ (see, e.g., \cite{Rubakov2}, Ch.21; \cite{Martin}, pp.15-20), can easily be expressed in terms of $\,\bl^\pr(\al),\, \bl^{\pp}(\al)$.\,\footnote{A good exercise is to express our parameters in terms of the standard ones and so to estimate them.}}

 \section{Brief summary and outlook}
 The main purpose of this paper is to advance a unified theory of dark matter, dark energy and inflation first formulated in 2008. Naturally, it combines new proposals and results with some old ones. Here we summarise only new ideas. Our minimal affine extension of the GR has geodesics coinciding with the pseudo Riemannian ones, up to parameterisations. It predicts a `\emph{sterile}' massive vecton and depends on \emph{two dimensional constants} $\La$ and $m^2$, which can be measured in the limit of small vecton velocity, $\da \rightarrow 0$ (neglecting gravity corrections). In the gauge dual to the standard Schwarzschild gauge, i.e., $\al(\tau)+\ga(\tau)=0$, the velocity has the upper bound: $\da^2<v^2_{\textrm{max}}\equiv\La/\ka$. When $\da^2\,\rightarrow\,v^2_{\textrm{max}}$
 the vecton kinetic energy becomes infinite, and in this sense the vecton is \emph{`relativistic'}, unlike the standard scalar inflaton. The linearized vecton theory is generally more similar to the scalaron models, its main distinction is the internal anisotropy. For this reason we study general anisotropic scalaron models with fairly general potentials $\bv(\al)$ and show how the isotropy may emerge during inflationary expansion. A similar mechanism must work in the linear vecton model.\footnote{In Section~\textbf{7.3}, an anisotropic vecton toy model is solved by the $\al$-approach invented for the scalaron theory. The complete linear vecton theory will be considered in another publication.}

 The main properties of the general scalaron model are described by three characteristic functions of $\al$: $\rbl, \wbl, \Theta$. All this functions are independent of $k$ and $\sig$ by definition, but $\wbl=\chi^2/6$, $\Theta=\Theta _{\chi}=-[\chi^2+\bl^\pr(\al)]$ when $k=\sig=0$. \emph{These  functions are better suited for qualitative and quantitative description of cosmological scenarios} than the potential $\bv(\al)$ or the exact solutions thus being more convenient  for constructing physical models. This is best illustrated by considering  inflationary phenomena. The general expressions for the potential $\bv(\al)$ and the generalized Hubble function $h(\al)$ in terms of our characteristic function are presented in Section~\textbf{5.3}.\footnote{In a less elegant form they were first derived in \cite{ATFnew}. During inflation, our $h(\al)$ changes very slowly but near the exit it may change very fast. Similar considerations can be used in analysis of bouncing scenarios.} The general analysis of Sections~\textbf{5.1-5.3} can be applied to concrete models to find realistic potentials and scenarios of exit from inflation and for search of possible new observable effects as well. The precise meaning of this proposal can be understood in the context of two-dimensional integrable models of gravity coupled to matter (see \cite{ATF2}-\cite{VDATF2}). The one-dimensional reduction of this theory has three sectors: cosmologies, static states (including black holes), and matter waves coupled to gravity. The one-dimensional sectors are independent, but in two and higher dimension they couple and it is possible to speak on waves of `matter' produced by perturbed cosmological evolutions.

 Analogously one can imagine that in realistic, non-integrable theory the end of inflation is strongly perturbed and couples to the matter-wave sector. This may result in \emph{creating spherical dark matter waves coupled to gravity instead of gravitational waves}. It looks that nobody tried to work in this direction, even in the frame of perturbed integrable models.

 More generally,  it is reasonable to suppose  that \emph{a sterile dark matter (coupled solely to gravity) may be the most substantial component of `matter' present at the very beginning of the Universe}. In this sense, the complex of problems and proposed solutions only touched upon in this paper looks deserving a wider attention.

 In conclusion, I would like to express my gratitude to the Bogoliubov Laboratory of JINR for permanent support of my work. I cannot be sure that I'll be able to complete the \emph{program announced in Introduction} and would greatly appreciate any questions, comments, and information related to the program and present work.

 \section{Appendix}
 
 \subsection{"Do it yourself\," sketches : $\wbl, \rbl, \Theta$ at possible exits}

 Initial intention of the author was to illustrate inflationary scenarios by drawing a few plausible graphics for the characteristic functions. But when you try to accomplish this task you find that their inflationary part is actually trivial and its essence is in a nontrivial exit from inflation as well as in  possible transitions to beginning of the `standard' cosmology. In fact, observable effects (if any?) must have their origin in this domain. In addition, in the author's opinion, the scalaron cosmology is not a realistic theory, which is expected to unify at least dark matter with dark energy and seriously take into account waves of dark matter.

 For all these reasons, I decided not to demonstrate my sketches of $\wbl$ and $\rbl$ for mentioned preferable cases \textbf{1}) and \textbf{2}) of Section \textbf{5.2} but give a comprehensible verbal description. The simplest case \textbf{2}) depends on fewer parameters but allows to draw my `ET-sketch'  (see discussions in Sections \textbf{5.2} - \textbf{5.3}). The case \textbf{1}) is much richer but, at the moment, they both are mainly illustrative and do not deserve serious discussion and comparison. I hope that interested readers (if any?) will find much better applications of sections \textbf{5.1-5.3}.

 The qualitative behavior of $\Theta(\al)$ corresponding to behavior of $\wbl(\al)$ defined by equations (\ref{mom11i}), (\ref{mom11k}), (\ref{mom20b}) is interesting by itself. The $\Theta(\al)$  is positive for $\al<\al_m $, negative for $\al > \al_m\,$, zero at $\al_m$. It has maximum $\Theta_+$ ($\al_i < \Theta_+ <\al_m$) and minimum $\Theta_-\,$ ($\al_m < \Theta_- <\al_j$). Its derivative $\Theta^\pr$ monotonically grows in the first and third intervals, and decreases in the second one. Note that $\Theta_+\,$ and $\Theta_-$ explicitly depend on $\,\bl^{\pp}(\al)<0$.

 \subsection{Solution of the anisotropy equation}
 
 Here we briefly discuss estimating $\sig(\rho)$ using  Eq.(\ref{mom11}) in the weak anisotropy limit. Recalling that $\bv(\rho)$ are rational function we estimate asymptotic behaviour of the isotropic approximation and then derive asymptotic of anisotropic corrections. When $k=0$, there exists the unique isotropic solution for $C_z=0$, otherwise we neglect the anisotropic solutions $\sig^\pr=(C_z/x(\rho)^{\half}$ decreasing like $\exp{(-3\rho)}$. Supposing that $\sig^\pr(\rho)\equiv\om(\rho)$ is small in the asymptotic region we can derive $\om(\rho)$ by iterations of the exact equation (\ref{mom8}). It is easily  transformed into\footnote{From now on we assume $C_z=0$, neglecting the correction to $\om$ of the order $\sim e^{-3\rho}$.}.
 \be
 [\,\om(\rho)\sqrt{x(\rho)}]^{\,\pr}\,=\,k\,e^{4\rho-2\sig}\,/
 \sqrt{x(\rho)}\,,
 \label{iso1}
 \ee
 that is immediately integrated (recall that $x(\rho)$ implicitly depends on $\sig$):
 \be
 \om(\rho)=\sig^\pr(\rho) = \frac{k}{\sqrt{x(\rho)}}\int \frac{e^{4\rho-2\sig}}{\sqrt{x(\rho)}}\,.
 \label{iso2}
 \ee
 Remembering that $x(\rho)\sim e^{6\rho}$ it is easy to see that $\om(\rho)\sim e^{-2\rho}$ and the dependence of r.h.s on $\sig$
 can be in first approximation neglected. Using the results of the main text we find:
 \be
 x(\rho) =  e^{6\rho}\,\bv(\rho)\,[1+\Sig_1(\rho)+ ke^{-2(\rho+\sig)}] + O(e^{-4\rho})\,,
 \label{iso3}
 \ee
 which gives the main asymptotic approximation for $\sig(\rho)$. 
 To simplify (\ref{iso2}) apply integration by parts using (\ref{iso3}) while writing the correct exponential behavior for the integrand,
 \be
 \int \frac{e^{4\rho-2\sig}}{\sqrt{x(\rho)}}\,=\,  \frac{e^{\rho-2\sig}\,\sqrt{6\,\bv^{-1}(\rho)}} {\sqrt{1+\Sig_1+ke^{-2(\rho+\sig)}}}\,-\,\int e^{\rho}\, \frac{d}{d\rho}\,\biggl(\frac{e^{-2\sig}\, \sqrt{6\,\bv^{-1}}}{\sqrt{1+\Sig_1+...}}\biggr)\,.
 \label{iso4}
 \ee

 The final result can thus be written in the simple and intuitively clear form:
 \be
 \sig(\rho) = -6k \int_\rho^\infty \frac{e^{-2\rho}\, [\bv(\rho)]^{-1}}{1+\Sigma_1(\rho) +ke^{-2\rho}}\,+\,o(e^{-2\rho}),
 \label{iso5}
 \ee
 where the integration limits correspond to the choice $C_z=0$. From here one can return to the second-order equation for $\sig$ (linear in derivatives) and solve it asymptotically by iterations. The first one is given by Eq,(\ref{iso1}). This result `finalizes' the requirement of asymptotic hierarchy $\rbl(\rho)\sim O(1)$, $\,\kbl(\rho)\sim O(e^{-2\rho})$, $\,\kbl\,\sig^\pr(\rho)\sim O(e^{-4\rho})$. The hierarchy very seriously constrains the potential $\bv$ from below. In fact, we must require that $\bv>v_0>0$, where $v_0$ cannot be theoretically estimated more precisely.  When $k=0$, the anisotropy can be  produced by the term $\sig^\pr = C_z/\sqrt{x(\rho)}$, so that the restriction on the potential is the same but its decreasing is even faster.\footnote{When $k=0$, the anisotropy $\sig(\rho)$ is independent of the potential and may be neglected by taking $C_z =0$. The restrictions on the potential are defined solely by a considered scenarios as discussed in the main text.} Special scenarios like the considered inflationary models (especially with exits) may lead to additional restrictions on possible potentials.

 \subsection{The simplest cosmology with vecton}
 
 Here, we consider special anisotropic cosmology with $\dal=\prbe=\prga=0$: 
 \be
   \cL_c\,=\,e^{2\beta}\bigl[e^{-\gamma}(\da^2\,- 2\dbe^2\,) - e^\ga(2\La+m^2 a^2)\bigr]- 6k e^{\gamma}\,.
   \label{ap2}
  \ee
 Denoting $\xi \equiv \dbe$, $b\equiv \da$, $v_0 \equiv 2\La$, $v(a) \equiv v_0+m^2 a^2$, we find the constraint
 \be
 e^{2\beta}\bigl[e^{-\gamma}(b^2\,- 2\xi^2\,) + e^\ga v(a)\bigr]+
  6k e^{\gamma} = 0\,,
 \label{ap3}
 \ee
 which is quadratic in momenta: $(p_a,\,p_\beta)\equiv 2e^{2\beta-\ga} (b\,,  -2\xi)$. Equations of motion are
 \be
 2\db + 2(2\xi-\dga)b + e^{2\ga}v^\pr(a) =0\,,
 \label{ap4}
 \ee 
 \be
 2\dxi + 4\xi^2 - 2\xi\dga - 2[v(a) + 3k e^{-2\beta}]e^{2\ga} = 0\,.
 \label{ap5}
 \ee
 With definitions $d/dt \equiv \xi d/d{\beta}$, $\xi\equiv \xi(\beta)$, $a\equiv a(\beta)$, $b \equiv b(\beta)$, $v(a)\equiv \bv(\beta)$, Eqs.(\ref{ap4})-(\ref{ap5}) are
 \be
 (b^2)^\pr + 2(2-\ga^\pr)b^2 + \bv^\pr(\beta)\,e^{2\ga}=0\,,
 \label{ap6}
 \ee
 \be
 (\xi^2)^\pr + 2(2-\ga^\pr)\xi^2-2\,[\bv(\beta)+3k e^{-2\beta}]\,e^{2\ga}= 0\,,
 \label{ap7}
 \ee
 where the dynamical variables depend on $\beta$ and we have taken into account that
 \be
 v^\pr(a)\frac{b}{\xi}\,=\, \frac{dv}{d\beta}\,\frac{d\beta}{da}\,\frac{da}{dt}\,\frac{dt}{d\beta} \,=\, \bv^\pr(\beta)\,. 
 \label{ap8}
 \ee

 Defining the new dynamical variables by $y(\beta)\equiv b^2\, e^{4\beta-2\ga}$, $x(\beta)\equiv\,\xi^2 e^{4\beta-2\ga}$ we find the gauge invariant equations and constraint similar to Eqs.(\ref{mom7}-\ref{mom9}) but there is no $\sig$-equation:
 \be
 x^\pr - 2e^{4\beta}\,\bv(\beta) +6k e^{2\beta} = 0\,, \quad (a); \qquad y^\pr + \bv^\pr(\beta)\,e^{4\beta} = 0\,,\quad (b)\,.
 \label{ap9}
 \ee
 \be
 2x(\beta) = y(\beta) + V(\beta) +6\,k\,e^{2\beta}\,, \quad  \textrm{where} \qquad V(\beta) \equiv e^{4\beta}\,\bv(\beta)\,.
 \label{ap10}
 \ee
 Their exact solution is also similar (\ref{mom10}) with  $\sig\equiv\,0$
 \be
 2x = 4I(\beta) + 6k e^{2\beta},\quad y = 4I(\beta) - V(\beta)\,;\qquad I(\beta) \equiv \, \int V(\beta)+C
 \label{ap11}
 \ee
 Evidently, $\bv(\beta)$ play the role of the potential $\bv(\al)$, and if we choose for it any analytic expression we will explicitly derive $\xi(\beta)=\dbe(t)$ and find $t(\beta)$ by one integration.  This allows to derive (not quite explicitly) $\beta(t)$ and $a(t)$ thus solving the model. Similarly to what has been done with solving the isotropic scalaron model and finding potentials $\bv(\al)$, one can construct inflationary potentials in this extremely anisotropic model. It is sufficient to define our main characteristic of cosmology
 \be
 \tchi^2(\beta) \equiv y(\beta)/2x(\beta) = 1 - V(\beta)/4I(\beta)\,, \quad \textrm{for}\quad k=0\,,
 \label{ap12}
 \ee
 which satisfy equation similar to equation (\ref{mom11dw}) for $\wbl(\al)$ equivalent to $\hchi^2(\al)$ when $k=0$:
 \be
 (\tchi^2)^\pr = - (\,1 - \tchi^2)\,(\,4\tchi^2 + \bl^{\,\pr})\, \equiv (\,1 - \tchi^2)\,\,\tThe(\beta)\,,\qquad \tchi^2 < 1\,. 
 \label{ap13}
 \ee
 Positivity of $\tThe(\beta)$ ensures the positivity of $(\tchi^2)^\pr$ and allows to find the conditions and relation defining the point of inflection on the curve $\tchi^2(\beta)$. This make it possible to construct inflationary potentials in this model. The second characteristic function, $\rbl(\beta)\equiv y(\beta)/V(\beta)$, is independent of $k$ and is expressed in terms of $\tchi^2(\beta)$ in Eq.(\ref{ap12}) when $k=0$ or $\beta\,\rightarrow +\infty$.

 It is not very difficult to apply almost all the constructions of the previous chapters to this vecton cosmology. Of course, any treatment of the models with the square-root kinetic terms requires some new ideas on characteristic functions. A special canonical transformation of the Hamiltonian proposed in  \cite{ATFp}-\cite{ATF14} might be the appropriate tool.

 \end{document}